\documentclass{article}
\usepackage{amsmath,epsfig,amssymb}
\usepackage{graphicx,color}
\usepackage{eufrak,mathrsfs}
\usepackage[mac]{inputenc}
\usepackage{array} 
\usepackage{upgreek,bm} 

\def\x{\boldsymbol{x}}
\def\m{\boldsymbol{m}}
\def\k{\boldsymbol{k}}
\def\u{\boldsymbol{u}}
\def\bw{\boldsymbol{\omega}}
\def\c{{\mathbf c}}
\def\Exp{{\mathrm e}}
\def\f{{\mathbf f}}
\def\w{{\mathbf w}}
\def\P{{\mathbf P}}
\def\F{{\mathbf F}}

\def\dual{\mathaccent'27}					
			
\newcommand{\sinc}{{\rm sinc}}

\newcommand{\mbb}{\mathbb}

\newcommand{\inner}[1]{\left\langle #1\right\rangle}

\DeclareSymbolFont{EUr}{U}{eur}{m}{n}
\DeclareSymbolFont{EUb}{U}{eur}{b}{n}
\DeclareMathSymbol{\varphi}{\mathord}{EUr}{"27}

\newtheorem{theorem}{Theorem}[section]

\newtheorem{proposition}[theorem]{Proposition}
\newtheorem{corollary}[theorem]{Corollary}

\begin{document}

\title{Construction of Hilbert Transform Pairs of Wavelet Bases and {G}abor-like Transforms}

\author{Kunal Narayan Chaudhury and Michael Unser\thanks{Corresponding Author: Kunal~Narayan~Chaudhury. The authors are with the Biomedical Imaging Group, Ecole Polytechnique Fédérale de Lausanne (EPFL),  Station-17, CH-1015 Lausanne VD, Switzerland. Fax: +41 21 693 37 01, e-mail: \{kunal.chaudhury, michael.unser\}@epfl.ch. This work was supported by the Swiss National Science Foundation under grant 200020-109415.}}

\maketitle

\begin{abstract}
We propose a novel method for constructing Hilbert transform (HT) pairs of wavelet bases based on a fundamental approximation-theoretic characterization of scaling functions---the B-spline factorization theorem. In particular, starting from  well-localized scaling functions, we construct HT pairs of biorthogonal wavelet bases of $\mathrm{L}^2(\mbb{R})$ by relating the corresponding wavelet filters via a discrete form of the continuous HT filter. As a concrete application of this methodology, we identify HT pairs of spline wavelets of a specific flavor, which are then combined to realize a family of complex wavelets that resemble the optimally-localized Gabor function for sufficiently large orders.

	 Analytic wavelets, derived from the complexification of HT wavelet pairs, exhibit a one-sided spectrum. Based on the tensor-product of such analytic wavelets, and, in effect, by appropriately combining four separable biorthogonal wavelet bases of $\mathrm{L}^2(\mbb{R}^2)$, we then discuss a methodology for constructing $2$D directional-selective complex wavelets. In particular, analogous to the HT correspondence between the components of the $1$D counterpart, we relate the real and imaginary components of these complex wavelets using a multi-dimensional extension of the HT---the directional HT.  Next, we construct a family of complex spline wavelets that resemble the directional Gabor functions proposed by Daugman. Finally, we present an efficient FFT-based filterbank algorithm for implementing the associated complex wavelet transform. 
\end{abstract}

\section{INTRODUCTION}
\label{sec:intro}

The dual-tree complex wavelet transform (DT-$\mbb{C}$WT) is a recent enhancement of the conventional  discrete wavelet transform (DWT) that has gained increasing popularity as a signal processing tool. The transform was  originally introduced by Kingsbury \cite{kingsbury1,kingsbury2} to circumvent the shift-variance of the decimated DWT, and involved two DWT channels in parallel with the corresponding wavelets forming a quadrature pair. In particular, Kingsbury realized the quadrature relation by interpolating the lowpass filters of one DWT ``mid-way'' between the lowpass filters of the other DWT. Moreover, based on appropriate combinations of separable wavelets, he extended the dual-tree construction to two-dimensions, where the corresponding transform, besides improving on the shift-invariance of the $2$D DWT, exhibits better direction selectivity as well. There is now good evidence that the transform tends to perform better than its real counterpart in a variety of applications such as such as deconvolution \cite{deconv}, denoising \cite{denoising}, and texture analysis \cite{texture}.

	 The crucial observation that the dual-tree wavelets involved in Kingsbury's construction form an approximate HT pair was made by Selesnick \cite{selesnick,selesnick_sobolev}. He also demonstrated that a particular phase relation between the lowpass (refinement) filters of the two channels resulted in the desired HT correspondence. This link consequently transposed the problem of designing different flavors of dual-tree wavelets to that of identifying new HT pairs of wavelets. Indeed, following this remarkable connection, several new paradigms and extensions have been proposed: design of HT pairs of biorthogonal wavelet bases \cite{ozkar}, alternative frameworks for complex non-redundant transforms \cite{CWT1}, and the M-band extension \cite{pasquet}, to name a few. 

\subsection{Motivation}
		
	 The deployment of complex signal representations for the determination of instantaneous amplitude and frequency is classical \cite{gabor, AMFM2}. Gabor and Ville \cite{gabor,ville} proposed to unambiguously define them using the concept of the \textit{analytic signal}---a unique complex-valued signal representation specified using the HT.  Specifically, the analytic signal $s_a(x)=s(x)+j\mathcal{H}s(x)$ corresponding to a real-valued signal $s(x)$ ($\mathcal H$ denotes the HT operator), was used to stipulate the instantaneous amplitude and phase via the polar representation $s_a(x)=A(x)\Exp^{j \phi(x)}$. In particular, this representation allows one to retrieve the time-varying amplitude and frequency of an AM-FM signal of the form $s(x)=A(x)\cos\left(2\pi \int_0^x \nu(\tau) d\tau+ \xi_0\right)$ via the estimates $A_{\mathrm{est}}(x)=|s_a(x)|$ and $\nu_{\mathrm{est}}(x)=(2\pi)^{-1}d\phi(x)/dx$, assuming $A(x)$ to be slowly-varying compared to $\nu(x)$. The analytic signal has become an important complex-valued representation in signal processing, especially in applications such as phase and frequency modulation, speech recognition and processing of seismic data. These concepts have also been transposed to the multi-dimensional setting: the local frequency has been used as a measure of local signal scale; structures such as lines and edges  have been distinguished using the local phase; and the local amplitude and phase have been used for edge detection and for texture and fingerprint analysis \cite{bovik}.
	
	The advantage of viewing the dual-tree wavelets as a HT pair is that we can make a direct connection with the formalism of analytic signals. Indeed, if we transpose the above concept to the wavelet domain and consider the input signal to be locally of the AM-FM form, we obtain a response where the local energy of the signal is encoded in the magnitude of the wavelet coefficients, while the relative displacement is captured by the phase. In fact, this turns out to be the fundamental reason for the superiority of the DT-$\mbb{C}$WT over conventional real-valued transforms whose response is necessarily oscillating.

\subsection{Our Contribution}
	
		In this contribution, we invoke the B-spline factorization theorem \cite{projection_biorthogonal}---a fundamental spectral factorization result---along with certain fractional B-spline calculus \cite{FS}, to construct HT pairs of biorthogonal wavelets from well-localized scaling functions. In particular, we do so by relating the corresponding wavelets filters via a discrete version of the continuous HT filter. 
		
		Next, we identify a family of analytic spline wavelets, of increasing vanishing moments and regularity, that asymptotically converge to Gabor-like functions \cite{gabor}. As far as the implementation is concerned, unlike Kingsbury's scheme that uses different filters for different stages (often with filter-swapping between the dual-trees), our implementation uses the same set of filters at all stages of the filterbank decomposition. Notably, we use an appropriate pair of projection filters for coherent signal analysis which, in turn, allows us to identify a discrete counterpart of the analytic wavelet---the so-called \textit{analytic wavelet filter} that exhibits a one-sided spectrum. 
		
		The construction is then extended to two-dimensions through appropriate tensor-products of the one-dimensional analytic wavelets. In particular, we construct a family of directional complex wavelets that resemble the directional Gabor functions proposed by Daugman \cite{Daugman} for sufficiently large orders. Moreover, we also relate the real and imaginary components of the complex wavelets using the directional HT---a multidimensional extension of the HT---that provides further insight into the directional-selectivity of the dual-tree wavelets.

\subsection{Organization of the Paper}

	  We begin by recalling certain fundamental definitions and properties pertaining to the HT and the fractional B-splines in $\S$\ref{II}. We characterize the action of the HT operator on B-splines in $\S$\ref{III}, which, along with the B-spline factorization theorem, is used to propose a formalism for constructing HT pairs of biorthogonal wavelet bases in $\S$\ref{IV}. The implementation aspects are discussed in $\S$\ref{V}. As a concrete application, we construct the Gabor-like wavelets in $\S$\ref{VI}. In $\S$\ref{VII},  directional complex wavelets are constructed by appropriately combining the wavelets corresponding to certain separable multiresolution analyses; the highlight of this section is the construction of $2$D Gabor-like spline wavelets. The implementation aspects of the corresponding $2$D Gabor-like transform are provided in $\S$\ref{VIII}, before concluding with $\S$\ref{IX}.

\section{PRELIMINARIES}
\label{II}

	We begin by introducing specific operators and functions that play a major role in the sequel followed by a discussion of their relevant properties. In what follows, we use $\hat{f}(\bw)=\int_{\mbb{R}^d} f(\x) \mathrm{e}^{-j\x^T \bw} \ d\x$ to denote the Fourier transform of a function $f(\x)$ on $\mbb{R}^d \ (d \geqslant 1)$, with $\x^T \bw$ being the usual inner-product on $\mbb{R}^d$. We also frequently use the notations $f(\cdot-\boldsymbol{s})$ and $f(\lambda \cdot)$, corresponding to some $\boldsymbol{s}$ in $\mbb{R}^d$ and $\lambda >0$, to denote the function obtained by translating (resp. dilating) $f(\x)$ by $\boldsymbol{s}$ (resp. $\lambda$). We denote the Kronecker-delta sequence by $\delta[n]$: its value is $1$ at $n=0$, and is zero at all other integers.   	
	
\subsection{Hilbert Transform and Wavelets}
	
	The Hilbert transform, that generalizes the notion of the quadrature transformation $\cos(\omega_0x)\mapsto \sin(\omega_0x)$ beyond pure sinusoids \cite{Bracewel}, forms the cornerstone of this paper. From a signal-processing perspective, the HT can be interpreted as a filtering operation in which the amplitude of the frequency components is left unchanged, while their phase is altered by $\pm \pi/2$  depending on the sign of the frequency. 

	Mathematically, the HT of a sufficiently well-behaved function is defined using a singular integral transform \cite{harmonic_analysis,Stein}. However, in the context of finite-energy signals, it admits a particularly straightforward formulation based on the Fourier transform on $\mathrm{L}^2(\mbb{R})$. In particular, the Hilbert transform on $\mathrm{L}^2(\mbb{R})$ is characterized by the equivalence 
\begin{equation}
\label{hilbert_def}
\mathcal{H}f(x) \stackrel{\mathcal{F}}{\longleftrightarrow} -j \mathrm{sign}(\omega) \widehat{f}(\omega)
\end{equation}
where the multiplier $\mathrm{sign}(\omega)$ is defined as $\omega/|\omega|$ for non-zero $\omega$, and as zero at $\omega=0$.

	Based on the above definition\footnote{The definition can also be extended to tempered distributions such as the Dirac delta and the sinusoid \cite[\S 2.5]{harmonic_analysis}.}, and the properties of the Fourier transform on $\mathrm{L}^2(\mbb{R})$, the following properties of the HT can be readily derived:
\begin{itemize}
\item   \textit{Linearity and Translation-invariance}: It is a linear and translation-invariant operator; that is, it acts as a convolution operator.

\item   \textit{Dilation-Invariance}: It commutes with dilations: $\mathcal{H}\{f(\lambda \cdot)\}(x)=(\mathcal{H}f)(\lambda x)$for all $\lambda >0$.

\item  \textit{Anti-Symmetry}: It anti-commutes with the flip operation $f^T(x)=f(-x)$, so that $(\mathcal{H}f^T)(x)=-(\mathcal{H}f)^{T}(x)$; thus the HT of a symmetric function is necessarily anti-symmetric.

\item   \textit{Unitary (Isometric) Nature}: It acts as a unitary operator on $\mathrm{L}^2(\mbb{R})$, so that $\inner{\mathcal{H}f,\mathcal{H}g}=\inner{f,g}$ for all  $f$ and $g$, where $\langle \cdot,\cdot \rangle$ denotes the usual inner-product on $\mathrm{L}^2(\mbb{R})$. Equivalently, this means that the inverse HT operator is given by its adjoint: $\mathcal{H}^{-1}=\mathcal{H}^{*}$. 
\end{itemize}

	It is well-known that HT of a wavelet is also a wavelet. The implication of the simultaneous invariance to dilations and translations is that the HT of a dilated-translated wavelet is a wavelet,  dilated and translated by the same amount:  $\mathcal{H} \{\psi (\lambda x - s)\}= (\mathcal{H}\psi)(\lambda x - s)$. Moreover, an immediate consequence of the unitary property is that the HT operator maps a basis into a basis: if $\{\psi_n\}$ form a (Riesz) wavelet basis of $\mathrm{L}^2(\mbb{R})$, then so does $\{\mathcal{H}\psi_n\}$.  It even preserves biorthogonal wavelet bases of $\mathrm{L}^2(\mbb{R})$: if $\{\psi_n\}$ and $\{\tilde\psi_m\}$ form a biorthogonal wavelet basis of $\mathrm{L}^2(\mbb{R})$, satisfying the duality criteria $\langle \psi_n,\tilde \psi_m \rangle=\delta[m-n]$, then using the same unitary property, we have
\begin{equation}
\langle \mathcal{H}\psi_n,\mathcal{H}\tilde\psi_m \rangle=\langle \psi_n,\tilde\psi_m \rangle=\delta[m-n]
\end{equation}
so that $\{\mathcal{H}\psi_n\}$ and $\{\mathcal{H}\tilde\psi_m\}$ form a biorthogonal wavelet basis of $\mathrm{L}^2(\mbb{R})$ as well. It is exactly the above invariance properties that make the construction of HT pair of wavelet bases of $\mathrm{L}^2(\mbb{R})$ feasible.

	Unfortunately, the HT exhibits certain inherent pathologies in the context of multiresolution analyses and wavelets. The impulse response of the HT, $\mathcal{H}\delta(x)=1/\pi x$ (in the sense of distributions), clearly indicates the non-local nature of the operator.  This has two serious implications: $(i)$ the HT of a compactly-supported scaling function/wavelet is no longer of finite support; $(ii)$ the HT-transformed function has a $O(1/|x|)$-decay in general, and hence is not integrable; and $(iii)$ the (anti-symmetric) HT suppresses the dc-component of symmetric scaling functions that is essential for fulfilling the partition-of-unity criterion. Therefore, the HT of a scaling function is not  a valid scaling function, and cannot be used to specify a multiresolution analysis in the sense of Mallat and Meyer \cite{mallat,meyer}. 

	Next, we recall the notion of an analytic signal that generalizes the phasor transformation transformation $\cos(\omega_0x) \mapsto  \exp(j\omega_0x)$ to finite-energy signals using the HT as the quadrature transformation. In general, the analytic signal $f_a(x)$ associated with a real-valued signal $f(x)$ is defined as the complex-valued signal   
\begin{equation}
f_a(x) = f(x)+j\mathcal{H}f(x).
\end{equation}
In particular, $f_a(x)=\exp(j\omega_0x)$ when $f(x)=\cos(\omega_0x)$. Importantly, note that the Fourier transform of the analytic signal evaluates to $\widehat{f_a}(\omega)=\left(1+\mathrm{sign}(\omega)\right) \widehat{f}(\omega)$, so that $\widehat{f_a}(\omega)$ vanishes for all negative frequencies. It is exactly this one-sided spectrum that makes the analytic signal particularly interesting in signal processing \cite{gabor}; we exploit this property for constructing directional wavelets in \S\ref{VII}.

\subsection{Fractional B-spline Multiresolution}
\label{valid_scaling} 

 	The family of fractional B-splines \cite{alpha-tau}---fractional extensions of the polynomial B-splines---will play a key role in the sequel. In particular, we recall that the fractional B-spline $\beta_{\tau}^{\alpha}(x)$, corresponding to a degree $\alpha \in \mathbb{R}^{+}_{0}$ and a shift $\tau \in \mathbb{R}$, is specified by its Fourier transform 
\begin{equation}
\label{spline_def}
\beta_{\tau}^{\alpha}(x) \stackrel{\mathcal{F}}{\longleftrightarrow}  \left( \frac{1-\Exp^{-j\omega}}{j\omega}\right)^{\frac{\alpha+1}{2}+\tau} \left( \frac{1-\mathrm{e}^{j\omega}}{-j\omega}\right)^{\frac{\alpha+1}{2}-\tau}. 
\end{equation}  
The parameters  $\alpha$ and $\tau$ control the width and the average group delay of the scaling function respectively. In particular, when $\tau=(\alpha+1)/2$, the fractional B-spline $\beta_{\tau}^{\alpha}(x)$ corresponds to the causal B-spline $\beta_{+}^{\alpha}(x)$ defined in \cite{FS}. The fractional B-splines, in general, do not have a compact support (except for integer degrees); however, their $O(1/|x|^{\alpha+2})$ decay ensures their inclusion in $\mathrm{L}^1(\mathbb{R}) \cap \mathrm{L}^2(\mbb{R})$. Another relevant property that will be invoked frequently is that the shift $\tau$ influences only the phase of the Fourier transform; that is, $|\widehat{\beta}_{\tau}^{\alpha}(\omega)|$ is independent of $\tau$.

	The fundamental role played by fractional B-splines in this paper is, however, based on the fact they satisfy certain admissibility criteria \cite{FS,alpha-tau} needed to generate a valid multiresolution of $\mathrm{L}^2(\mbb{R})$: \newline
$\mathbf{(C1)}$ The approximation space $\mathcal{V}_0=\mathrm{span}_{\ell^2} \{\beta_{\tau}^{\alpha}(\cdot-k)\}_{k \in \mathbb{Z}}$ admits a stable Riesz basis. \newline
$\mathbf{(C2)}$ There exists an integrable sequence $h_{\tau}^{\alpha} [k]$ (refinement filter) such that the two-scale relation  
 \begin{equation}
 \frac{1}{2} \beta_{\tau}^{\alpha} \left(\frac{x}{2}\right)= \sum_{k \in \mbb{Z}} h_{\tau}^{\alpha} [k] \beta_{\tau}^{\alpha} (x-k)
 \end{equation}
holds. In particular, the transfer function of the refinement filter is specified by  
 \begin{equation}
H_{\tau}^{\alpha}(\mathrm{e}^{j\omega})= \frac{1}{2^{\alpha+1}} \left(1+\mathrm{e}^{j\omega}\right)^{\frac{\alpha+1}{2}-\tau} \left(1+\Exp^{-j\omega}\right)^{\frac{\alpha+1}{2}+\tau}. 
\end{equation} 
$\mathbf{(C3)}$ Partition of unity: The integer-translates of $\beta_{\tau}^{\alpha}(x)$ can reproduce the unity function. \newline

     We briefly discuss the significance of these admissibility conditions. The criterion $\mathbf{(C1)}$ ensures a stable and unique representation of functions in $\mathcal{V}_0$ using coefficients from $\mathrm{\ell}^2(\mbb{Z})$; equivalently, this also signifies that the transfer function of the autocorrelation (Gram) filter, $A^{\alpha}(\mathrm{e}^{j\omega})=\sum_{k \in \mathbb{\mathbb{Z}}} |\widehat{\beta}_{\tau}^{\alpha}(\omega+2\pi k)|^2$, is uniformly bounded from above, and away from zero \cite{WTSP}. On the other hand, $\mathbf{(C2)}$ implies the inclusion of $\beta_{\tau}^{\alpha}(x/2)$ in $\mathcal{V}_0$, which, in turn, allows one to define a hierarchical embedding of approximation spaces $\{\mathcal{V}_j\}_{j \in \mbb{Z}}$ that  is key to the multiresolution structure of the associated wavelet transform. Finally, the technical condition $\mathbf{(C3)}$ ensures that the multiresolution $\{\mathcal{V}_j\}$ is dense in $\mathrm{L}^2(\mbb{R})$: arbitrarily close approximations of functions in $\mathrm{L}^2(\mbb{R})$ can be achieved using elements from $\{\mathcal{V}_j\}$.

\section{HILBERT TRANSFORM AND B-SPLINES}
\label{III}

 It turns out that the action of the HT on B-splines can be effectively characterized in terms of certain fractional finite-difference (FD) operators. In particular, corresponding to an order $\alpha \in \mathbb{R}_{0}^{+}$ and shift $\tau \in \mathbb{R}$, we consider the operator $\Delta^{\alpha}_{\tau}$ defined on $\mathrm{L}^2(\mbb{R})$ by 
 \begin{equation}
\Delta^{\alpha}_{\tau}f(x)  \stackrel{\mathcal{F}}{\longleftrightarrow} D^{\alpha}_{\tau}(\mathrm{e}^{j\omega}) \widehat{f}(\omega),
 \end{equation}
where $D^{\alpha}_{\tau}(\mathrm{e}^{j\omega})=\left(1-\Exp^{-j\omega}\right)^{\frac{\alpha}{2}+\tau} \left(1-\mathrm{e}^{j\omega}\right)^{\frac{\alpha}{2}-\tau}$. 

	One recovers the conventional $n$-th order FD operator by setting $\alpha=n$ and $\tau=n/2$. Since the operator has a periodic frequency response, one can associate with it a digital filter $d^{\alpha}_{\tau}[k]$ through the correspondence $\Delta^{\alpha}_{\tau}f(x)=\sum d^{\alpha}_{\tau}[k] f(x-k).$ The FD operator that is especially relevant for our purpose is the zeroth-order operator $\Delta^0_{-1/2}$ (henceforth, we simply denote it by $\Delta$). The corresponding frequency response $D(\mathrm{e}^{j\omega})=D^0_{-1/2}(\mathrm{e}^{j\omega})$ reduces\footnote{We specify the fractional power of a complex number $z$ by $z^{\gamma}=|z|^{\gamma}\Exp^{j\gamma \arg(z)}$ corresponding to the principal argument $|\arg(z)| <\pi$. On this principal branch, the identity $(z_1 z_2)^{\gamma}=z_1^{\gamma}z_2^{\gamma} $ holds only if $\arg(z_1)+ \arg(z_1) \in (-\pi,\pi)$ \cite[Chapter 3]{Stein_ComplexAnalysis}.} to
\begin{equation}
D(\mathrm{e}^{j\omega})=-j  \mathrm{sign}(\omega) \Exp^{-j\frac{\omega}{2}} \quad \mathrm{for}  \  \omega \in (-\pi,\pi),      
 \end{equation}
signifying that $D(\mathrm{e}^{j\omega})$ is in $\mathrm{L}^2((-\pi,\pi))$; the corresponding filter coefficients $d[k]=d^{0}_{-1/2}[k]$ in $\ell^2(\mbb{Z})$ are then specified\footnote{The inverse Fourier transform over the principal period $(-\pi,\pi)$ is invoked.} by 
\begin{eqnarray}
d[k]&=&\frac{1}{2\pi} \int_{-\pi}^{\pi} D(\mathrm{e}^{j\omega}) \Exp^{j k\omega} \mathrm{d}  \omega  \nonumber \\
&= &\frac{1}{2\pi} \int_{-\pi}^{\pi} -j  \mathrm{sign}(\omega) \Exp^{j(k+1/2)\omega} \mathrm{d}  \omega   \nonumber \\
&=& \frac{1}{\pi(k+1/2)}, \ k \in \mbb{Z}.
\end{eqnarray}
Thus, similar to the HT operator, $\Delta$ is also unitary, and the corresponding filter $d[k]$ can be interpreted as a discrete form of the continuous HT filter $1/\pi x$. In particular, we can relate the action of the HT on the B-splines solely in terms of this filter. Indeed, it can easily be seen that the Fourier transform of the B-spline can be factorized as
\begin{equation}
\widehat{\beta}_{\tau}^{\alpha}(\omega)= (j \omega)^{1/2} (-j\omega)^{-1/2}  D (\mathrm{e}^{j\omega}) \widehat{\beta}_{\tau+1/2}^{\alpha}(\omega),
\end{equation}
which, along with the identity $(j \omega)^{\frac{1}{2}}(-j\omega)^{-\frac{1}{2}}=j\mathrm{sign}(\omega)$, results in the equivalence
\begin{align}
\mathcal{H} \beta_{\tau}^{\alpha}(x) &\stackrel{\mathcal{F}}{\longleftrightarrow}  -j  \mathrm{sign}(\omega) \widehat{\beta}_{\tau}^{\alpha}(\omega) \nonumber \\
 &=  -j \ \mathrm{sign}(\omega)\cdot j\mathrm{sign}(\omega) D (\mathrm{e}^{j\omega}) \widehat{\beta}_{\tau+1/2}^{\alpha}(\omega) \nonumber \\
 &=  D (\mathrm{e}^{j\omega}) \widehat{\beta}_{\tau+1/2}^{\alpha}(\omega) \nonumber \\ 
 & \stackrel{\mathcal{F}}{\longleftrightarrow}    \Delta\beta_{\tau+1/2}^{\alpha}(x),
\end{align}
that establishes the desired result:
\vspace{0.05in}
\begin{proposition}
The HT of a fractional B-spline can be expressed as 
\begin{eqnarray}
\label{hilbert}
\mathcal{H}\beta_{\tau}^{\alpha}(x)= \sum_{k \in \mathbb{Z}} \frac{1}{\pi (k+ 1/2)}  \beta_{\tau+1/2}^{\alpha}(x-k)  .
\end{eqnarray} 
 \end{proposition}
 \vspace{0.05in}
In particular, the digital filter $d[k]$ acts as a unitary convolution operator on $\mathrm{L}^2(\mbb{R})$ when applied to functions, and as a discrete filter on $\mathrm{\ell}^2(\mathbb{Z})$ when applied to sequences. The theoretical difficulty with the HT stems from the fact that its frequency response has a singularity at $\omega=0,$ which results in a poor decay of the transformed output. The remarkable feature of \eqref{hilbert} is that we have been able to express the slowly decaying HT as a linear combination of the better-behaved B-splines. Specifically, the sequence $d[k]$ decays only as $O(1/|k|)$, whereas $\beta_{\tau+1/2}^{\alpha}(x)$ decays as $O(1/|x|^{\alpha+2})$. 

	Thus, by expressing the HT using shifted B-splines as in  \eqref{hilbert}, we have, in effect, moved the singularity onto the digital filter. In the sequel, we shall apply this filter to the wavelets where its effect is much more innocuous since $\widehat\psi(\omega)=0$ around the origin. \\

\textit{Half-Delay Filters}: As remarked earlier, the shift parameter $\tau$ only affects the phase of the Fourier transform of the fractional B-spline and the corresponding refinement filter \cite{alpha-tau}. 
In particular, based on the factorization
\begin{align*}
\label{filters}
H_{\tau+\frac{1}{2}}^{\alpha}(\mathrm{e}^{j\omega})
&= \frac{1}{2^{\alpha+1}} \left(1+\mathrm{e}^{j\omega}\right)^{\frac{\alpha+1}{2}-(\tau+\frac{1}{2})} \left(1+\Exp^{-j\omega}\right)^{\frac{\alpha+1}{2}+(\tau+\frac{1}{2})} \nonumber \\
&=  (1+\mathrm{e}^{j\omega})^{-1/2} (1+\Exp^{-j\omega})^{1/2} H_{\tau}^{\alpha}(\mathrm{e}^{j\omega})\nonumber \\
&= \Exp^{-j\frac{\omega}{2}} H_{\tau}^{\alpha}(\mathrm{e}^{j\omega}),  \ \mathrm{for}  \ \omega \in (-\pi,\pi),
\end{align*}
we arrive at the following result:
\vspace{0.05in}
\begin{proposition} 
The spline refinement filters $h^{\alpha}_{\tau}[k]$ and $h^{\alpha}_{\tau+1/2}[k]$ are ``half-sample'' shifted versions of one another in the sense that
\begin{equation}
\label{half-shift}
H_{\tau+1/2}^{\alpha}(\mathrm{e}^{j\omega})=\Exp^{-j\frac{\omega}{2}} H_{\tau}^{\alpha}(\mathrm{e}^{j\omega})   
\end{equation} 
for all $\omega$ in $(-\pi,\pi)$.
\end{proposition}
\vspace{0.05in}

	Indeed, if we consider the bandlimited function $h^{\alpha}_{\tau}(x)=\sum h^{\alpha}_{\tau}[k] \sinc(x-k)$ that satisfies the constraint $h^{\alpha}_{\tau}(x)|_{x=k}=h^{\alpha}_{\tau}[k]$, then we have, as a consequence of \eqref{half-shift}, the relation $h^{\alpha}_{\tau+1/2}[k]=h^{\alpha}_{\tau}(k-1/2)$: each filter provides the bandlimited interpolation of the other mid-way between its samples. 

Finally, we make a note of the fact that the above refinement filters can also be related through a \textit{conjugate-mirrored} version of the FD filter:
\begin{equation}
\label{filter_pair}
H_{\tau+1/2}^{\alpha}(\mathrm{e}^{j\omega})= D(-\Exp^{-j\omega}) H_{\tau}^{\alpha}(\mathrm{e}^{j\omega}).
\end{equation}

\section{HT PAIR OF WAVELET BASES}
\label{IV}

Before stating the main results, we recall the approximation-theoretic notion of \textit{approximation order}, and a fundamental spectral factorization result involving B-splines. \\

\textit{Approximation Order}: Scaling functions play a fundamental role in wavelet theory. The technical criteria for a valid scaling function was discussed earlier in the context of B-splines (cf. $\S$\ref{valid_scaling}). Next we recall the fundamental notion of order for a scaling function that characterizes its approximation power \cite{projection_biorthogonal}. A scaling function $\varphi(x)$ is said to have an approximation order $\gamma$ if and only if there exists a positive constant $C$ such that for all elements of the Sobolev space $\mathrm{W}_2^{\gamma}(\mbb{R})$, of order $\gamma$, we have the estimate
\begin{equation}
||f- P_a f|| \leqslant C a^{\gamma} ||\partial^{\gamma}f||. 
\end{equation}
Here $P_a$ denotes the projection operator from $\mathrm{W}_2^{\gamma}(\mbb{R})$ onto the approximation subspace $\mathrm{span}_{\ell^2}\{\varphi(\cdot/a-k)\}_{k \in \mbb{Z}}$, and $\partial^{\gamma}$ denotes the (distributional) derivative of order $\gamma$. In other words, the approximation order provides a characterization of the rate of decay of the approximation error for sufficiently regular functions as a function of the scale. 

	It turns out that, akin to their polynomial counterparts, the order of fractional B-splines is entirely controlled by their degree \cite{FS,alpha-tau}; in particular, we have $\gamma=\alpha+1$. Equivalently, this signifies that any polynomial of degree $\leqslant \lceil \alpha \rceil$ can be reproduced by the set $\{\beta_{\tau}^{\alpha}(\cdot-k)\}$, which is crucial for capturing the lowpass information in images is concerned.   \vspace{0.05in}

\textit{Characterizazion of Scaling Functions}: A fundamental result in wavelet theory is that it is always possible to express a valid scaling function as a convolution between an fractional B-spline and a distribution \cite {projection_biorthogonal}. The original result in \cite {projection_biorthogonal} involves causal B-splines; however, the result can readily be extended to the more general fractional B-splines since the shift parameter $\tau$ does not influence the order of the scaling function. Indeed, note the theorem in \cite {projection_biorthogonal} asserts that $H(\mathrm{e}^{j\omega})$ is the refinement filter  of a valid scaling function (cf. $\S$\ref{valid_scaling}) of order $\alpha+1$ if and only if it can be factorized as    
\begin{equation}
\label{basic}
H(\mathrm{e}^{j\omega})=\stackrel{\beta^{\alpha}_+  \mathrm{\ spline \ part}}{\overbrace{\Big(\frac{1+\Exp^{-j\omega}}{2}\Big)^{\alpha+1}}}  \stackrel{\mathrm{distributional \ part}}{\overbrace{Q(\mathrm{e}^{j\omega})}},
\end{equation}
where $Q(\mathrm{e}^{j\omega})$ is stable: $|Q(\mathrm{e}^{j\omega})|<C<+\infty$ for all $\omega$. Rewriting \eqref{basic} in terms of a $(\alpha,\tau)$ B-spline refinement filter, we then have the following equivalent representation:
\begin{equation}
\label{ext}
H(\mathrm{e}^{j\omega})=\stackrel{\beta^{\alpha}_{\tau} \mathrm{\ spline \ part}}{\overbrace{\Big(\frac{1+\Exp^{-j\omega}}{2}\Big)^{\frac{\alpha+1}{2}+\tau}  \Big(\frac{1+\mathrm{e}^{j\omega}}{2}\Big)^{\frac{\alpha+1}{2}-\tau}}}  \stackrel{\mathrm{distributional \ part}}{\overbrace{P(\mathrm{e}^{j\omega})}} 
\end{equation}
with $P(\mathrm{e}^{j\omega})= \Exp^{-j\omega\left(\frac{\alpha+1}{2}-\tau\right)}Q(\mathrm{e}^{j\omega})$for $\omega \in (-\pi,\pi)$. Note that $P(\mathrm{e}^{j\omega})$ is stable, with $|P(\mathrm{e}^{j\omega})|< C<+\infty$ for all $\omega$. That is, $H(\mathrm{e}^{j\omega})$ is the refinement filter of a valid scaling function of order $\alpha+1$ if and only if it admits a stable factorization as in \eqref{ext}. We then arrives at the following extension: 
\vspace{0.05in}
\begin{theorem} (\emph{B-spline Factorization})
\label{T1}
A valid scaling function $\varphi(x)$ is of order $\alpha+1$ if and only if its Fourier transform can be factorized as 
\begin{equation}
\widehat\varphi(\omega)=\widehat\beta_{\tau}^{\alpha}(\omega) \widehat\varphi_0(\omega)
\end{equation}
for some $\tau \in \mathbb{R}$, where $\widehat\varphi_0(\omega)$ is a function of $\omega$ that is bounded on every compact interval, and equals unity at the origin. 
\end{theorem}
\vspace{0.05in}

	In the signal domain, this corresponds to a well-defined convolution $\varphi(x)=(\beta_{\tau}^{\alpha}\ast \varphi_0)(x)$ between a B-spline and the tempered distribution $\varphi_0$. The crux of the above result is that it is the constituent B-spline that is solely responsible for the approximation property, and other desirable features of the scaling function \cite {projection_biorthogonal}.

\subsection{Construction of HT Pairs of Wavelets}

	 In what follows, we use the notation $f_{j,k}(x)$, corresponding to a function $f(x)$, and integers $j$ and $k$, to denote the (normalized) dilated-translated function $2^{j/2}f(2^jx-k)$. The HT of a wavelet is also a wavelet in a well-defined sense. In particular, if $\psi(x)$ is a wavelet whose dilations-translations $\{\psi_{j,k}\}$ form a Riesz basis of $\mathrm{L}^2(\mbb{R})$, then $\mathcal{H}\psi(x)$ is also a valid wavelet with $\{\mathcal{H}\psi_{j,k}\}$ constituting a Riesz basis of $\mathrm{L}^2(\mbb{R})$. As remarked earlier, this follows from the fundamental invariance properties of the HT. 
 
   We now establish a formalism for constructing the HT of a given wavelet $\psi(x)$. In particular, if $\varphi(x)$ be the associated scaling function, say of order $\alpha+1$, and $g[k]$ be the generating wavelet filter, then we have the relation
\begin{equation}
\psi(x/2)=\sum g[k] \varphi(x-k).
\end{equation}
Following Theorem \ref{T1}, let us factorize $\varphi(x)$ as
\begin{equation}
\varphi(x)=(\beta_{\tau}^{\alpha}\ast \varphi_0)(x)
\end{equation}
corresponding to some real $\tau$. Then, consider the scaling function $\varphi'(x)$, of the same order, specified by $\varphi'(x)=(\beta_{\tau+1/2}^{\alpha}\ast \varphi_0)(x)$. Let $\psi'(x)$ be any arbitrary wavelet, corresponding to the multiresolution analysis associated with $\varphi'(x)$, that is specified by $\psi'(x/2)=\sum g'[k] \varphi'(x-k)$. We then have the following necessary and sufficient condition for the desired HT correspondence in terms of the discrete HT filter $d[k]$ (see \S\ref{A1} for a proof):
\vspace{0.05in}
\begin{theorem} (\emph{HT Pair of Wavelets})
 \label{construction} 
The wavelets $\psi(x)$ and  $\psi'(x)$ have the correspondence $\psi'(x)=\mathcal{H}\psi(x)$ if and only if $g'[k]=(d\ast  g)[k]$. \end{theorem}
\vspace{0.05in}
Moreover, the construction has the following characteristics:
\begin{itemize}
\item  Both $\varphi(x)$ and $\varphi'(x)$ have the same Riesz bounds and the same decay,
\item The refinement filters $H(\mathrm{e}^{j\omega})$ and $H'(\mathrm{e}^{j\omega})$ corresponding to $\varphi(x)$ and $\varphi'(x)$ respectively, are related as
\begin{equation*}
H'(\mathrm{e}^{j\omega})=\Exp^{-j\frac{\omega}{2}} H(\mathrm{e}^{j\omega})
\end{equation*}
for all $\omega$ in $(-\pi,\pi)$.
\end{itemize}
	The equality of the Riesz bounds follows from the observation that the autocorrelation filters of $\varphi(x)$ and $\varphi'(x)$ are identical. Indeed, we have
\begin{equation*}
\begin{split}
 a[k]&= \langle \varphi, \varphi(\cdot-k) \rangle =(\beta_{0}^{2\alpha+1} \ast \varphi_0\ast\varphi^T_0)(k),   \\
 a'[k]&= \langle \varphi', \varphi'(\cdot-k) \rangle=(\beta_{0}^{2\alpha+1} \ast \varphi_0\ast\varphi^T_0)(k).
 \end{split}
 \end{equation*}
The assertion regarding the decay is based on the observation that both $\beta_{\tau}^{\alpha}(x)$ and $\beta_{\tau+1/2}^{\alpha}(x)$ have the same decay. Finally, using \eqref{half-shift} and \eqref{ext}, we can relate the transfer functions on $(-\pi,\pi)$ as follows  
 \begin{align*}
 H'(\mathrm{e}^{j\omega})&= H^{\alpha}_{\tau+1/2}(\mathrm{e}^{j\omega}) P(\mathrm{e}^{j\omega})\nonumber   \\ 
 &= \Exp^{-j\frac{\omega}{2}} H_{\tau}^{\alpha}(\mathrm{e}^{j\omega}) P(\mathrm{e}^{j\omega})   \nonumber  \\ 
 &=  \Exp^{-j\frac{\omega}{2}} H(\mathrm{e}^{j\omega}),     
 \end{align*}   
where $P(\mathrm{e}^{j\omega})$ denotes the transfer function of the filter associated with the distribution $\varphi_0$.  

 	\textit{Remark}: Note that although $\mathcal{H}\psi(x)$ is unique, the scaling function $\varphi'(x)$ and the corresponding filter $g'[k]$ generating $\mathcal{H}\psi(x)$ are by no means unique. For instance,  the particular choice $\varphi'(x)=\mathcal{H}\varphi(x)$ and $g' \equiv g$ is sufficient to ensure that $\psi'(x)=\mathcal{H}\psi(x)$. Moreover, if  $\varphi'(x)$ and $g'[k]$ generate the wavelet $\psi'(x/2)=\sum g'[k] \varphi'(x-k)$ such that $\psi'(x)=\mathcal{H}\psi(x)$, then so do $\varphi'_{\mathrm{eq}}(x)=\sum r[k] \varphi'(x-k)$ and $g'_{\mathrm{eq}}[k]=(g' \ast r_{\mathrm{inv}})[k]$. Here the filter $r[k]$ is such that $0<|\sum_k r[k] \Exp^{-j\omega k}| <+\infty$ for all $\omega$ so that the convolutional inverse $r_{\mathrm{inv}}[k]$ is well-defined. 
    
	The condition $g'[k]=(d\ast  g)[k]$ is both necessary and sufficient only for our preferred choice of the scaling function $\varphi'(x)=(\beta_{\tau+1/2}^{\alpha}\ast \varphi_0)(x)$. This particular choice of the scaling function against the more direct choice $\mathcal{H}\varphi(x)$ is justified on the following grounds: 
\begin{itemize}
\item The function $\varphi'(x)$ is well-localized with better decay properties than $\mathcal{H}\varphi(x)$; the latter is not even integrable in general (e.g., the Harr scaling function), 

\item  The scaling function $\varphi'(x)$ satisfies the partition-of-unity requirement, whereas $\mathcal{H}\varphi(x)$ is not a valid scaling function since $\widehat{\mathcal{H}\varphi}(0)$ is not necessarily unity. For example, if $\varphi(x)$ is symmetric and $\mathcal{H}\varphi(x)$ is integrable, then we have $\widehat{\mathcal{H}\varphi}(0)=\int (\mathcal{H}\varphi)(x) dx=0$ following the fact that $\mathcal{H}\varphi(x)$ is anti-symmetric. 
\end{itemize}
 
\subsection{HT Pairs of Biorthogonal Wavelets}
  	
  	 A biorthogonal wavelet basis of $\mathrm{L}^2(\mbb{R})$, corresponding to the dual-primal scaling function pair $(\varphi,\tilde \varphi)$ of order $( N+1, \tilde N+1)$, involves the nested multiresolution
\begin{equation*}
\{0\}  \subset \cdots \subset  \mathcal{V}_{-1} \subset \mathcal{V}_0 \subset \mathcal{V}_1 \subset   \cdots   \subset \mathrm{L}^2(\mbb{R}),
\end{equation*}
and its dual 
\begin{equation*}
\{0\}  \subset \cdots \subset  \mathcal{\tilde V}_{-1} \subset \mathcal{\tilde V}_0 \subset \mathcal{\tilde V}_1\subset   \cdots \subset \mathrm{L}^2(\mbb{R}),
\end{equation*}
where the approximation subspace $\mathcal{V}_j$ (resp. $\mathcal{\tilde V}_j$) is generated by the translations of $\varphi_{j,0}(x)$ (resp. $\tilde\varphi_{j,0}(x)$) \cite{WTSP}. Let $(\psi, \tilde \psi)$ be the wavelets associated with these multiresolutions, which, along with their dilated-translated copies, encode the residual signal---the difference of the signal approximations in successive subspaces. In particular, the wavelet $\psi_{j,0}(x)$ (resp. $\tilde{\psi}_{j,0}(x)$) and its translates span the complementary space $\mathcal{W}_j=\mathcal{V}_{j} \ominus \mathcal{V}_{j-1}$ (resp. $\tilde{\mathcal{W}_j}=\tilde{\mathcal{V}}_{j}  \ominus \tilde{\mathcal{V}}_{j-1}$). The crucial aspect of the construction is that the dilated-translated ensemble $\psi_{j,k}(x)$ and $\psi_{j',k'}(x)$ form a dual basis of $\mathrm{L}^2(\mbb{R})$, i.e., they satisfy the biorthogonality criteria $\langle  \psi_{j,k}, \tilde \psi_{j',k'}\rangle=\delta[j-j',k-k']$. The expansion of a finite-energy signal $f(x)$ in terms of this biorthogonal basis is then given by 
\begin{equation}
 f(x)=\sum_{(j,k)\in  \mbb{Z}^2} \langle f, \tilde \psi_{j,k}\rangle \psi_{j,k}(x).
\end{equation}
In other words, the wavelets $\{\tilde \psi_{j,k}(x)\}$ and $\{\psi_{j,k}(x)\}$, interpreted as the analysis and synthesis wavelets respectively, together constitute a biorthognal wavelet basis of $\mathrm{L}^2(\mbb{R})$. 

	In particular, let $\tilde\varphi(x)$ and $\varphi(x)$ be the scaling functions, of order $\tilde N+1$ and $N+1$ respectively, associated with a given biorthogonal wavelet basis, with associated wavelets 
\begin{align*}
\tilde{\psi}(x/2)&=\sum \tilde g[k] \tilde \varphi(x-k), \ \psi(x/2)=\sum g[k] \varphi(x-k).
\end{align*}
Now, let $\tilde\varphi(x)=(\beta_{\tilde\tau}^{\tilde{N}}\ast \tilde{\varphi}_0)(x)$ and $\varphi(x)=(\beta_{\tau}^{N}\ast \varphi_0)(x)$ be the respective factorizations of $\tilde\varphi(x)$ and $\varphi(x)$. Consider the scaling functions $\tilde\varphi'(x)=(\beta_{\tilde\tau+1/2}^{\tilde N}\ast \tilde{\varphi}_0)(x)$ and $\varphi'(x)=(\beta_{ \tau+1/2}^{N}\ast \varphi_0)(x)$, with associated wavelets specified by
\begin{equation*}
\tilde{\psi}'(x/2)=\sum \tilde g'[k] \tilde{\varphi}'(x-k), \ \psi'(x/2)=\sum  g'[k]  \varphi'(x-k).
\end{equation*}
Then the following result comes as a direct consequence of Theorem \eqref{construction}. 
\vspace{0.05in}
\begin{corollary}(\emph{HT Pair of Biorthogonal Wavelets})
\label{corollary} 
The following are equivalent: 
\begin{itemize}
\item The primal and dual wavelets form HT pairs, $\tilde\psi'(x)=\mathcal{H}\tilde\psi(x)$ and $\psi'(x)=\mathcal{H}\psi(x)$, and $\{\tilde{\psi}'_{j,k}(x)\}$ and $\{\psi'_{j',k'}(x)\}$ together constitute a biorthogonal wavelet basis of $\mathrm{L}^2(\mbb{R})$.
\item  The discrete HT correspondences $\tilde g'[k]=(d\ast \tilde g)[k]$ and $g'[k]=(d\ast g)[k]$ hold.
\end{itemize}
\end{corollary}
\vspace{0.05in}
	The above construction also exhibits the following properties: 
\begin{itemize}
\item The two biorthogonal systems have the same order, and the same Riesz bounds.

\item  If the pair $(\tilde\varphi, \varphi)$ satisfy the biorthogonality relation, then so do $(\tilde\varphi', \varphi')$. Indeed, using the identity $\mathcal{H}\beta_{\tau+1/2}^{\alpha}(x)=-\Delta^0_{1/2}\beta_{\tau}^{\alpha}(x)$, we can express the inner-product $\langle \tilde\varphi',\varphi'(\cdot-k)\rangle$ as
\begin{align*}
 & \langle \beta_{\tilde\tau+1/2}^{\tilde N}\ast \tilde{\varphi}_0,(\beta_{ \tau+1/2}^{N}\ast \varphi_0)(\cdot-k) \rangle \nonumber \\
&= \langle \mathcal{H}\big(\beta_{\tilde\tau+1/2}^{\tilde N}\ast \tilde{\varphi}_0\big),\mathcal{H}\big(\beta_{ \tau+1/2}^{N}\ast \varphi_0\big)(\cdot-k)\rangle \nonumber \\
&=  \langle -\Delta^0_{1/2} \big(\beta_{\tilde\tau}^{\tilde N}\ast \tilde{\varphi}_0\big),-\Delta^0_{1/2}\big(\beta_{ \tau}^{N}\ast \varphi_0\big)(\cdot-k) \rangle   \nonumber \\
&= \langle \tilde\varphi,\varphi(\cdot-k) \rangle,
\end{align*} 
which establishes the assertion.

\item The lowpass filters on both the analysis and synthesis side are ``half-sample'' shifted versions of one another, and are related via the modulation of the discrete HT filter: 
\begin{eqnarray*}
\label{H_relation}
\tilde H(z^{-1})&=& D(-z^{-1}) \tilde{H}'(z^{-1}),  \nonumber \\
H'(z)&=&  D(-z^{-1}) H(z). 
\end{eqnarray*}
In particular, the filter is ``half-sample" delayed on the analysis side, whereas on the synthesis side the filter has a ``half-sample" advance.

\item The highpass filters on both the analysis and synthesis side are related through the FD filter as
\begin{eqnarray*}
\label{G_relation}  
\tilde G(z^{-1})&=& D(z) \tilde G'(z^{-1}), \nonumber \\
G'(z)&=& D(z) G(z).
\end{eqnarray*}

\item  If the analysis and synthesis filters of the original biorthogonal system satisfy the PR conditions 
\begin{eqnarray*}
\label{PR1}
&&G(z^{-1}) \tilde G(z) + H(z^{-1})\tilde H(z)=1, \nonumber \\
&&G(z^{-1})\tilde G(-z) + H(z^{-1})\tilde H(-z)=0,
\end{eqnarray*}
then so do the filters of the HT pair. Indeed, since $D(z) D(z^{-1})=1$, we have
\begin{align*}
G'(z^{-1}) \tilde G'(z) &+ H'(z^{-1})\tilde H'(z) \nonumber \\
&= D(z^{-1}) D(z)G(z^{-1}) \tilde G'(z)  \nonumber \\
&+  D(-z)  D(-z^{-1}) H(z^{-1})\tilde H(z) \nonumber \\
&=G(z^{-1}) \tilde G(z) + H(z^{-1})\tilde H(z).
\end{align*}
Similarly,  $G'(z^{-1})\tilde G'(-z) + H'(z^{-1})\tilde H'(-z)=0.$ 
\end{itemize}
Note that above properties relate to a common theme: the unitary nature of the operators $\mathcal{H}$ and $\Delta$ involved in the wavelet and the filterbank construction, respectively.

\section{$1$D IMPLEMENTATION}
\label{V}

\textit{Signal Pre-filtering}: In order to implement the DT-$\mbb{C}$WT, we need to employ two parallel wavelet decompositions corresponding to the wavelets $\psi(x)$ and $\psi'(x)$. Moreover, to have a coherent signal analysis---same input applied to both wavelet branches---we need to project the input signal $f(x)$ separately onto $V(\varphi)$ and $V(\varphi')$ before applying the respective DWTs. In particular, given a finite-energy input signal $f(x)$, we consider its orthogonal projection $f_0(x)=\sum c_0[k] \varphi(x-k)$ onto the space $V(\varphi)$. The $J$-level wavelet decomposition of the signal $f_0(x)$ is subsequently given by
\begin{equation}
f_0(x)=\sum_{k \in \mbb{Z}} c_J[k] \varphi_{J,k}(x) \ + \sum_{1 \leqslant j \leqslant J , \ k \in \mbb{Z}} d_j[k] \psi_{j,k}(x),
\end{equation}
where the wavelet coefficients $d_j[k]$, and the coarse approximation coefficients $c_J[k]$ are recursively derived from the projection coefficients $c_0[k]$ using Mallat's filterbank algorithm \cite{mallat}.

	However, in practice one has access only to the discrete samples of the input signal $f(x)$; let $\{f[k]\}_{k \in \mathbb{Z}}$ be such (uniform) signal samples. It turns out that by assuming the input signal $f(x)$ to bandlimited, a particularly simple digital filtering algorithm for computing the projection coefficients is obtained:
\begin{equation}
\label{prefilter_formula}
c_0[k]=(f \ast p)[k],
\end{equation}
where the frequency response $P(\mathrm{e}^{j\omega})$ of the digital filter $p[k]$ is uniquely specified by the restriction $P(\mathrm{e}^{j\omega})=\widehat{\dual\varphi}(\omega)$ for $\omega \in (-\pi,\pi)$ (derivation details in \S\ref{A3}). 

	As for the second branch, the input signal is projected onto the corresponding approximation space $V(\varphi')$: the same type of pre-filtering is applied with an appropriate modification of the frequency response, i.e., $\widehat{\dual\varphi'}(\omega)$ is used instead of $ \widehat{\dual\varphi}(\omega)$. To implement  the filters for finite input signals, we use a FFT-based algorithm, similar to the one used in \cite{FFT} for implementing the DWT filters.   \\

\textit{Analysis \& Reconstruction}: To simplify the notation, we shall henceforth use matrix notation to represent the linear transformations associated with the discrete DT-$\mbb{C}$WT. For instance, corresponding to an input signal $\f \in \mbb{R}^N$, the least-square projections are specified by $\c^L_0=\P \f$ and $\mathbf{c'}_0^L=\P' \f$, where $\P$ and $\P'$ are the $N \times N$ circulant matrices corresponding to the two pre-filters. 
	
	Let $(\tilde h,\tilde g, h,g)$ and $(\tilde h',\tilde g', h',g')$ be the set of perfect-reconstruction filters associated with the biorthogonal systems $(\tilde\varphi,\tilde\psi,\varphi,\psi)$ and $(\tilde\varphi',\tilde\psi',\varphi',\psi')$ respectively. The lowpass $\{(\c^L_i, \mathbf{c'}^L_i)\}$ and the highpass $\{(\c^H_i, \mathbf{c'}^H_i)\}$ subbands at successive levels $i=1,\ldots, J$ are then given by the recursive filterbank decompositions:
\begin{align}
\label{dwt}
\c^L_i &=\F_{\tilde h}  \c^L_{i-1},  \quad \  \ \c^H_i=\F_{\tilde g} \c^L_{i-1} \nonumber  \\
\mathbf{c'}^L_i &=\F_{\tilde h'} \mathbf{c'}^L_{i-1}, \quad   \mathbf{c'}^H_i=\F_{\tilde g'} \mathbf{c'}^L_{i-1}, 
\end{align}
where $\F_{\tilde h}$ and $\F_{\tilde g}$ (resp. $\F_{\tilde h'}$ and $\F_{\tilde g'}$) denote the composition of the downsampling matrix and the DWT matrix representing the lowpass and highpass analysis filters of the first (resp. second) channel. The complex wavelet subbands $\w_1,\ldots,\w_J$ are then specified by $\w_i=\c^H_i+j\mathbf{c'}^H_i$. In fact, the analysis can be summarized by the single frame operation 
\begin{equation}
T: \f \mapsto \big(\c^L_J, \mathbf{c'}^L_J, \w_1,\ldots,\w_J\big).
\end{equation}
from a lower-dimensional space to a higher-dimensional space: $\dim(T\f) > \dim(\f)$.

  	In several signal processing applications (e.g., denoising) one also needs to perform an inverse transform, that is, reconstruct the denoised signal from the processed complex wavelet coefficients. Since $T$ is realized through the concatenation of bases, it is injective:  $Tf=T \acute f$ only if $f=\acute f$; however, as result of the redundancy, $T$ exhibits non-unique left-inverses. In our case, we use a simple left-inverse:
\begin{equation*}
T^{\dagger} : \big(\c^L_J, \mathbf{c'}^L_J, \w_1,\ldots,\w_J\big) \mapsto \f=\frac{1}{2}(\P^{-1} \c^L_0+\P'^{-1} \mathbf{c'}^L_0),
\end{equation*}
where $(\c^L_0,\mathbf{c'}^L_0)$ are obtained via the recursion
\begin{align}
\label{trivial_inverse}
\c^L_{i}&=\F_{h}   \c^L_{i+1}+\F_{g} \mathfrak{Re}(\w_{i+1}), \nonumber \\
 \mathbf{c'}^L_{i}&=\F_{h'}   \mathbf{c'}^L_{i+1}+\F_{g'} \mathfrak{Im}(\w_{i+1}),
\end{align}
for $i=J-1,\ldots,0$ ($\mathfrak{Re}(z)$ and $\mathfrak{Im}(z)$ denote the real and imaginary components of $z$ respectively). Here $\F_{h}$ and $\F_{g}$ (resp. $\F_{h'}$ and $\F_{g'}$) represent the composition of the DWT matrix corresponding to the lowpass and highpass synthesis filters of the first (resp. second) channel and the upsampling matrix. In short, the above inversion operation essentially amounts to inverting the two parallel transforms and averaging the inverses.   

\begin{figure}
\centering
\includegraphics[width=0.70\linewidth]{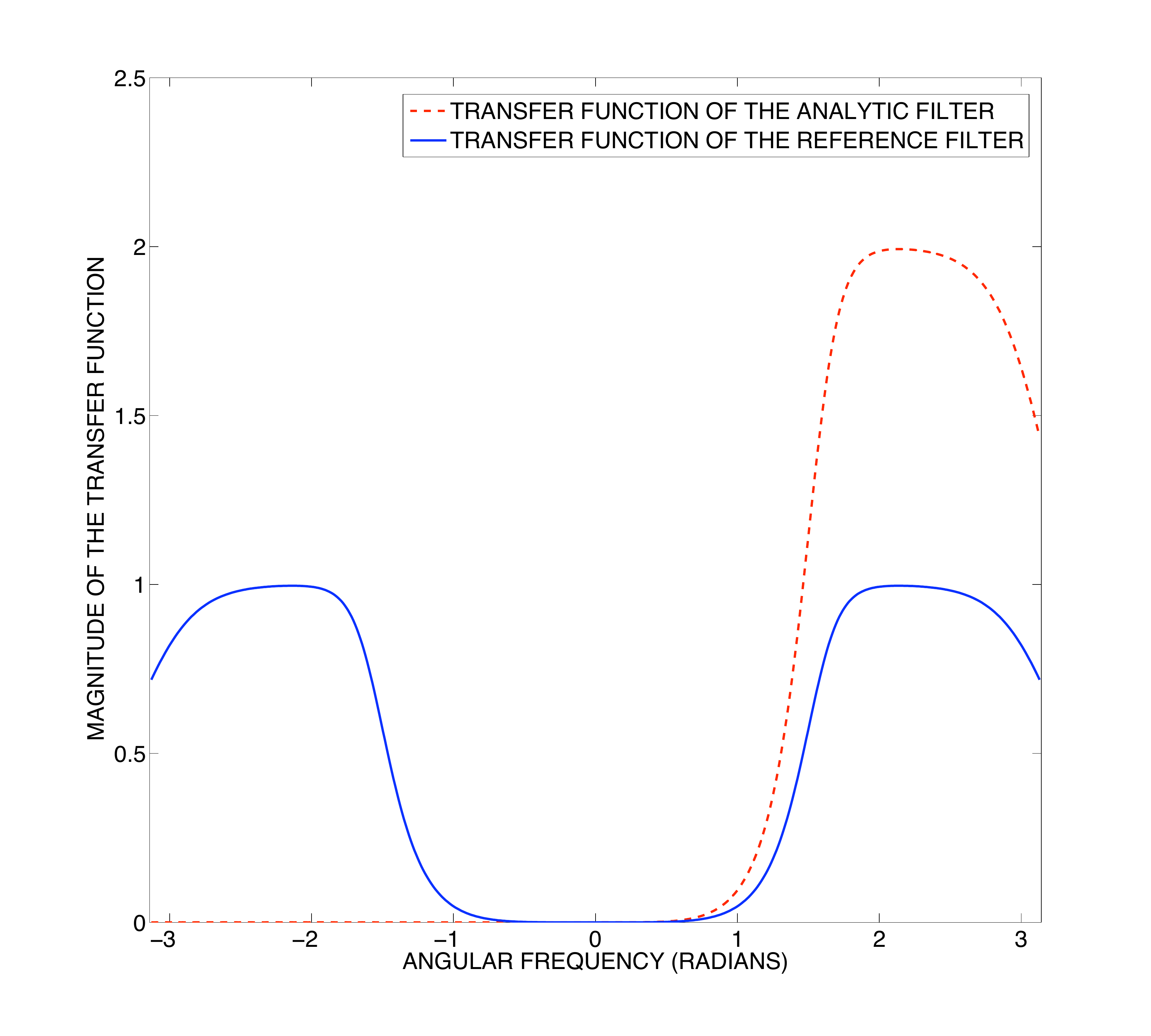}  
\caption{Transfer function of the \textit{analytic} wavelet filter $P_a(\Exp^{j\omega})$.}	
\label{One_sided_response}
\end{figure}

\textit{Remark}:	The role played by the two projection filters $P(\mathrm{e}^{j\omega})$ and $P'(\mathrm{e}^{j\omega})$ is critical as far as the issue of \textit{analyticity} is concerned. Note that, while the analytic wavelet has an exact one-sided Fourier transform (by construction), the corresponding complex wavelet filter $\tilde{G}(\mathrm{e}^{j\omega})+j\tilde{G}'(\mathrm{e}^{j\omega})$ does not inherit this property naturally; it is only the combination of the projection and wavelet filters, $P_a(\mathrm{e}^{j\omega})=P(\mathrm{e}^{j\omega}) \tilde{G}(\mathrm{e}^{j\omega}) + j P'(\mathrm{e}^{j\omega}) \tilde{G}'(\mathrm{e}^{j\omega}),$ that exhibits this property: $P_a(\mathrm{e}^{j\omega})=0,$ for $\omega \in (-\pi,0]$. Figure \ref{One_sided_response} shows the one-sided magnitude response of the filter.

\section{GABOR-LIKE WAVELETS}
\label{VI}

	 The ``quantum law'' for information---the principle that the joint time-frequency domain of signals is quantized, and that the joint time-frequency support of signals always exceed a certain minimal area---was enunciated in signal theory by Dennis Gabor \cite{gabor}. He also identified the fact that the family of Gaussian-modulated complex exponentials (and their translates) provide the best trade-off in the sense of Heisenberg's uncertainty principle. 
	
	 The canonical Gabor transform analyzes a signal using the set of ``optimally-localized'' Gabor atoms:  
\begin{equation}
\label{gabor_basis}
g_{m,n}(x)=\frac{1}{\sqrt{2\pi}T_1} \exp\Big({-\frac{1}{2T_1^2}(x-mT)}\Big)  \Exp^{j n\Omega (x-mT)}
\end{equation}
generated via the modulations-translations of a Gaussian-modulated complex exponential pulse \cite{gabor,gabor_expansion}. In particular, this paradigm involves the analysis of a finite-energy signal $f(x)$ using the discrete sequence of projections $c_{m,n}=\inner{f,g_{m,n}}$ corresponding to different modulations and translations $(m,n) \in \mathbb{Z}^2$. 

	Note that the Gabor atoms have a fixed size (specified by the width $T_1$ of the Gaussian window), and hence the associated transform essentially results in a ``fixed-window'' analysis of the signal. Moreover, the analysis functions $\{g_{m,n}(x)\}$ form a frame \cite{WTSP} and not a basis of $\mathrm{L}^2(\mathbb{R})$; consequently, the reconstruction process involving the dual frame is often computationally expensive and/or unstable \cite{gabor_expansion}. 

\subsection{Analytic Gabor-like Wavelets}
 
   As a concrete application of the ideas developed in \S\ref{IV}, we now construct a family of analytic spline wavelets that asymptotically converge to Gabor-like functions. In particular, we consider the family of semi-orthogonal B-spline wavelets that are better localized in space than their orthonormal counterparts, and that exhibit remarkable joint time-frequency localization properties \cite{convergence}. 
   
   	In particular, consider the multiresolution in \S\ref{valid_scaling}, generated by the fractional B-spline $\beta^{\alpha}_{\tau}(x)$. The transfer function of the wavelet filter that generates the so-called B-spline wavelet \cite{semi-ortho} associated with this multiresolution is specified by 
\begin{equation}
\label{semi_filter}
G_{\tau}^{\alpha}(\mathrm{e}^{j\omega})=\mathrm{e}^{j\omega} A^{\alpha}(-\mathrm{e}^{j\omega})H_{\tau}^{\alpha}(-\Exp^{-j\omega}).
\end{equation}
We denote the wavelet by $\psi_{\tau}^{\alpha}(x)$. The dual multiresolution (resp. wavelet) is specified by the unique dual-spline function $\dual{\beta}_{\tau}^{\alpha}(x)$ (resp. dual wavelet, denoted by $\tilde{\psi}_{\tau}^{\alpha}(x)$). 

	Following Corollary \ref{corollary}, it can be shown (proof provided in \S\ref{A2}) that the family of B-spline wavelets $\{\psi_{\tau}^{\alpha}(x)\}_{\tau \in \mbb{R}}$, of a fixed order $\alpha$, and their duals are closed with respect to the HT:
\vspace{0.05in}
\begin{proposition}\emph{(HT Pair of B-spline Wavelets)}
\label{prop_main}
The HT of a B-spline (resp. dual-spline) wavelet is a B-spline (resp. dual-spline) wavelet of same order, but with a different shift:
\begin{eqnarray}
\label{main}
\mathcal{H}\psi_{\tau}^{\alpha}(x)=\psi_{\tau+1/2}^{\alpha}(x), \nonumber \\
\mathcal{H}\tilde{\psi}_{\tau}^{\alpha}(x)=\tilde{\psi}_{\tau+1/2}^{\alpha}(x).
\end{eqnarray}
\end{proposition}
\vspace{0.05in}
	
	The importance of this result is that  it allows us to identify the analytic B-spline wavelet of degree $\alpha$ and shift $\tau$: 
\begin{equation}
\Psi^{\alpha}_{\tau}(x)=\psi_{\tau}^{\alpha}(x)+j\psi_{\tau+1/2}^{\alpha}(x).
\end{equation}
In the sequel (cf. \S \ref{2Dgabor}), we shall make particular use of this analytic spline wavelet. We would, however, like to highlight a different aspect: the remarkable fact that the wavelet $\Psi^{\alpha}_{\tau}(x)$ resembles the celebrated Gabor functions for sufficiently large $\alpha$. Indeed, it was shown in \cite{convergence} that the B-spline wavelets asymptotically converge to the real part of the Gabor function; by appropriately modifying the proof in \cite{convergence}, the following asymptotic convergence can established:
\begin{equation}
\label{thiery}
\psi_{\tau}^{\alpha}(x) \underset{\alpha \rightarrow +\infty}{\sim} M \exp\left(-\frac{(x-1/2)^2}{2\sigma^2}\right) \cos\Big(\omega_0 x -\frac{\omega_0}{2}-\pi \tau\Big).
\end{equation}
where $M=2M_0^{\alpha+1} \Delta\omega_0/\sqrt{2 \pi (\alpha+1)}; \sigma=\sqrt{\alpha+1}/\Delta \omega_0,$ with $M_0=0.670, \omega_0=-5.142$ and $\Delta\omega_0=2.670$. We recall that the asymptotic notation $f_\alpha(x) \sim g_\alpha(x)$ signifies that $f_\alpha(x)/g_\alpha(x) \rightarrow 1$ as $\alpha \rightarrow +\infty$ for all $x$. Immediately, we have the following result: 
\vspace{0.05in}
\begin{proposition}
\label{gabor_1D}
\emph{(Gabor-like Wavelet)}
The complex B-spline wavelet $\Psi^{\alpha}_{\tau}(x)$ resembles the Gabor function for sufficiently large $\alpha$:
\begin{equation}
\label{asymp_formula}
\Psi^{\alpha}_{\tau}(x)  \sim M \exp\left(-\frac{(x-1/2)^2}{2\sigma^2}\right)\Exp^{j\left(\omega_0 x -\frac{\omega_0}{2}-\pi \tau\right)} . 
\end{equation}
\end{proposition}
\vspace{0.05in}
	
\begin{figure}
\centering 
\fbox{\includegraphics[width=70mm,height=90mm]{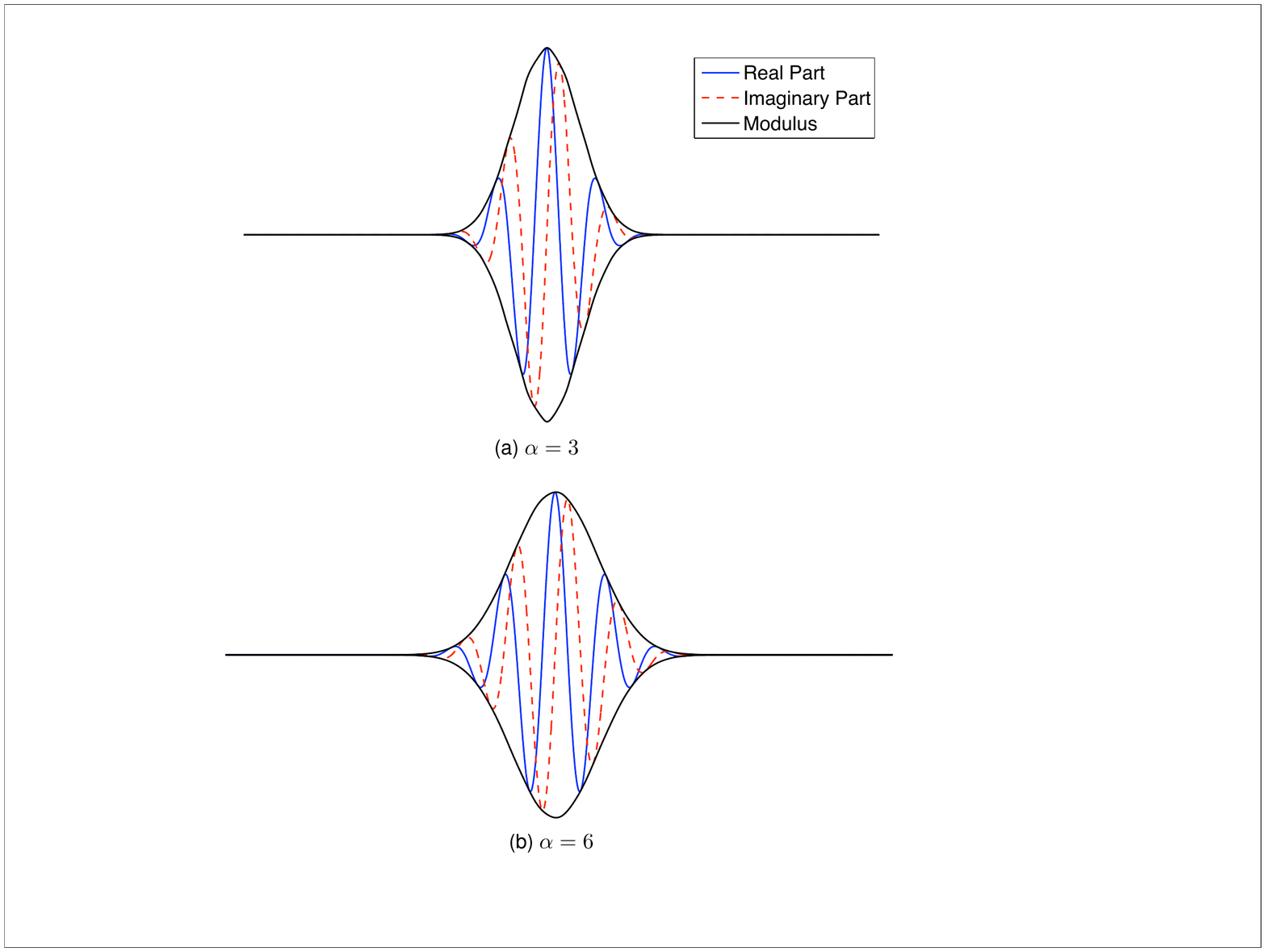}}  
\caption{HT pairs of B-spline wavelets. In either case, Blue (solid line): $\psi_{0}^{\alpha}(x)$, Red (broken line): $\psi_{1/2}^{\alpha}(x)$, Black (solid line):  $|\psi_0^{\alpha}(x)+\psi_{1/2}^{\alpha}(x)|$}
\label{Gabor-Wavelet}
\end{figure}

	The above convergence happens quite rapidly. For instance, we have observed that the joint time-frequency resolution of the complex cubic B-spline wavelet ($\alpha=3$) is already within 3\% of the limit specified by the uncertainty principle. Fig. \ref{Gabor-Wavelet} depicts the complex wavelets generated using HT pair of B-spline wavelets; the wavelets becomes more Gabor-like as the degree increases. Also shown in the figure is the magnitude envelope $|\Psi^{\alpha}_{\tau}(x)|$ of the complex wavelet which closely resembles the well-localized Gaussian window of the Gabor function. From a practical viewpoint, this means that one could use the non-redundant and numerically stable multiresolution spline  transforms to approximate the Gabor analysis.  

\vspace{0.05in}
	
\begin{figure}
\centering 
\fbox{\includegraphics[width=50mm,height=70mm]{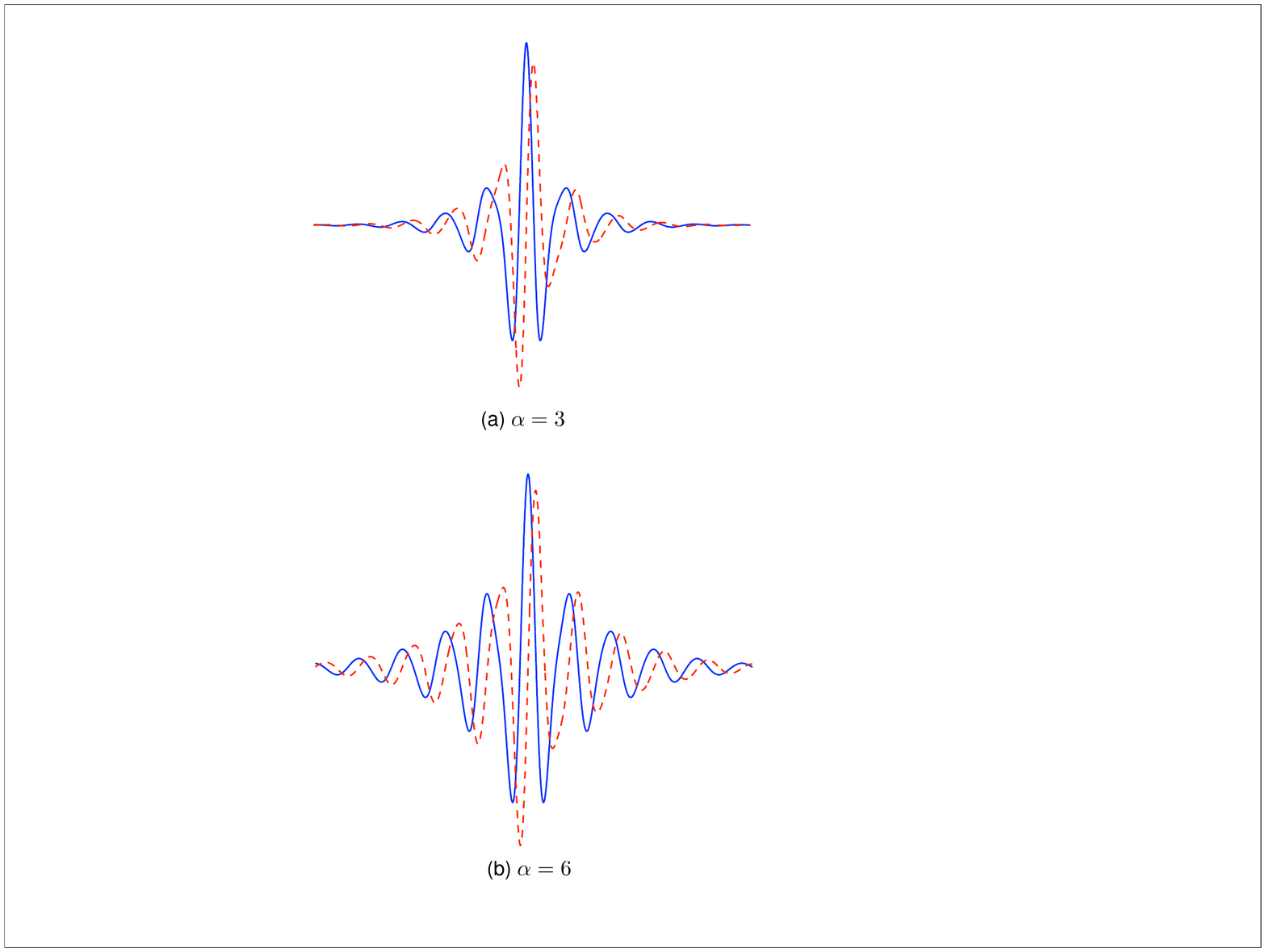}}  
\caption{HT pairs of dual-spline wavelets. In either case, Blue (solid line): $\psi_0^{\alpha}(x)$, Red (broken line): $\psi_{1/2}^{\alpha}(x)$.}
\label{primal_dual_pairs}
\end{figure}

\textit{Remark}: While the B-spline wavelets tend to be optimally localized in space, we have already observed that they are not orthogonal to their translates. The reconstruction therefore requires the use of some complementary dual functions. The flip side is that these dual-spline wavelets have a comparatively poor spatial localization, that deteriorates as the degree increases. This is evident in Fig. \ref{primal_dual_pairs}, which shows quadrature pairs of such wavelets of different degrees. However, we should emphasize that the dual (synthesis) wavelets have the same mathematical rate of decay as their analysis counterpart, and that the associated reconstruction algorithm is fast and numerically stable.

\subsection{Gabor-like Transform}
\label{impl}

 	The dual-tree Gabor-like transform is based on the analytic B-spline wavelet $\Psi^{\alpha}_0(x)=\psi^{\alpha}_0(x)+j \psi^{\alpha}_{1/2}(x)$, where the degree $\alpha$ is sufficiently large (the choice $\tau=0$ is arbitrary). The analysis and synthesis filters for the first and second channel are as specified below \cite{semi-ortho}:

\begin{itemize}
\item {First channel:
\begin{align}
\label{DWT1}
\tilde{H}(z)&=2^{-(\alpha+1)}(1+z)^{\frac{\alpha+1}{2}}(1+z^{-1})^{\frac{\alpha+1}{2}} , \nonumber  \\
\tilde{G}(z)&=z A^{\alpha}(-z)H(-z^{-1}),  \nonumber \\
H(z)&= \tilde{H}(z)A^{\alpha}(z)/A^{\alpha}(z^2),  \nonumber \\
G(z)&=\tilde{G}(z)/A^{\alpha}(z^2)A^{\alpha}(-z) .
\end{align}}

\item {Second channel:
\begin{align}
\label{DWT2}
\tilde{H}'(z)&=2^{-(\alpha+1)}(1+z)^{\frac{\alpha}{2}}(1+z^{-1})^{\frac{\alpha}{2}+1} , \nonumber \\
\tilde{G}'(z)&=z A^{\alpha}(-z) H'(-z^{-1}) , \nonumber \\
H'(z)&= \tilde{H'(z)}A^{\alpha}(z)/A^{\alpha}(z^2), \nonumber \\
G'(z)&=\tilde{G}'(z)/A^{\alpha}(z^2)^{\alpha}A(-z) .
\end{align}}
\end{itemize}

The DT-$\mbb{C}$WT corresponding to this Gabor-like wavelet would then result in the analysis of the input signal $f(x)$ in terms of the sequence of multiscale projections $\langle f, \sqrt {2^m} \Psi^{\alpha}_0(2^m \cdot -k)\rangle$ onto the (normalized) dilated-translated templates of the Gabor-like wavelet $\Psi^{\alpha}_0(x)$. Note that here the Gabor-like wavelet is used for analysis, whereas its dual is used for synthesis. The corresponding DWTs $(\F_{\tilde h}, \F_{\tilde g}, \F_{h}, \F_{g})$ and $(\F_{\tilde h'}, \F_{\tilde g'}, \F_{h'}, \F_{g'})$ are efficiently implemented using a practical FFT-based algorithm, outlined in \cite{FFT}. This method is exact despite the infinite support of the underlying wavelets, and achieves perfect-reconstruction up to a very high accuracy. The pre-filters, $\P$ and $\P'$, are also implemented in a similar fashion. An added advantage of the frequency domain implementation is that the execution time is independent of the order of the spatial filters. Moreover, the filters in \eqref{DWT1} need to be pre-computed once and for all in order to apply the transform to different signals (of a fixed length). 

\section{BIVARIATE EXTENSION}
\label{VII}

	Next, based on ideas similar to those of Kingsbury \cite{CTDWT}, we construct $2$D complex wavelets, and $2$D Gabor-like wavelets in particular, using a tensor-product approach. Moreover, we also relate the real and imaginary components of the complex wavelets using a multi-dimensional extension of the HT. 	\\  

	\textit{Separable Biorthogonal Wavelet Basis}: Biorthogonal wavelet bases of $\mathrm{L}^2(\mbb{R})$ can be combined to construct a biorthogonal wavelet basis of $\mathrm{L}^2(\mathbb{R}^2)$. The underlying principle used to construct such a basis using tensor-products is as follows \cite{WTSP}: 
\vspace{0.05in}
\begin{theorem}
\label{prop}
Let $(\psi_p,\tilde \psi_p)$ be the primal and dual wavelets of a biorthogonal wavelet basis of $\mathrm{L}^2(\mbb{R})$, with corresponding scaling functions $(\varphi_p,\tilde \varphi_p)$. Similarly, let $(\psi_q,\tilde \psi_q)$ constitute another biorthogonal wavelet basis with corresponding scaling functions $(\varphi_q,\tilde \varphi_q)$. Consider the following separable wavelets and their duals
\begin{alignat}{2}
\psi_1(\x)&=\varphi_p(x)\psi_q(y), & \hspace{8mm} \tilde\psi_1(\x)&=\tilde\varphi_p(x)\tilde\psi_q(y), \nonumber \\
\psi_2(\x)&=\psi_p(x)\varphi_q(y), & \hspace{8mm}  \tilde\psi_2(\x)&=\tilde\psi_p(x)\tilde\varphi_q(y), \nonumber \\
\psi_3(\x)&=\psi_p(x)\psi_q(y). & \hspace{8mm}    \tilde\psi_3(\x)&=\tilde\psi_p(x)\tilde\psi_q(y).
\end{alignat}  
Then the dilation-translations of $(\psi_1(\x),\psi_2(\x),\psi_3(\x))$ and $(\tilde{\psi}_1(\x),\tilde{\psi}_2(\x),\tilde{\psi}_3(\x))$ together constitute a biorthogonal wavelet basis of $\mathrm{L}^2(\mathbb{R}^2)$. \\
\end{theorem} 
\vspace{0.05in}
	 The functions $\psi_1(\x),\psi_2(\x)$ and $\psi_3(\x)$ are popularly referred to as the `low-high' (LH), `high-low' (HL) and `high-high' (HH) wavelets, respectively, to emphasize the directions along which the lowpass scaling function and the highpass wavelet operate (here $\x=(x,y)$ denote the spatial coordinates). Note that the primal and dual approximation spaces for the above construction are $V(\varphi_p) \otimes V(\varphi_q)$ and $V(\tilde\varphi_p) \otimes V(\tilde\varphi_q)$ respectively, where $V(\varphi_p) \otimes V(\varphi_q)$ denotes the subspace $\mathrm{span} \{\varphi_p(\cdot-m)\varphi_q(\cdot-n)\}_{(m,n) \in \mbb{Z}^2}$.

\subsection{Wavelet Construction}
\label{directionality}

\begin{figure}
\centering
\includegraphics[width=0.65\linewidth]{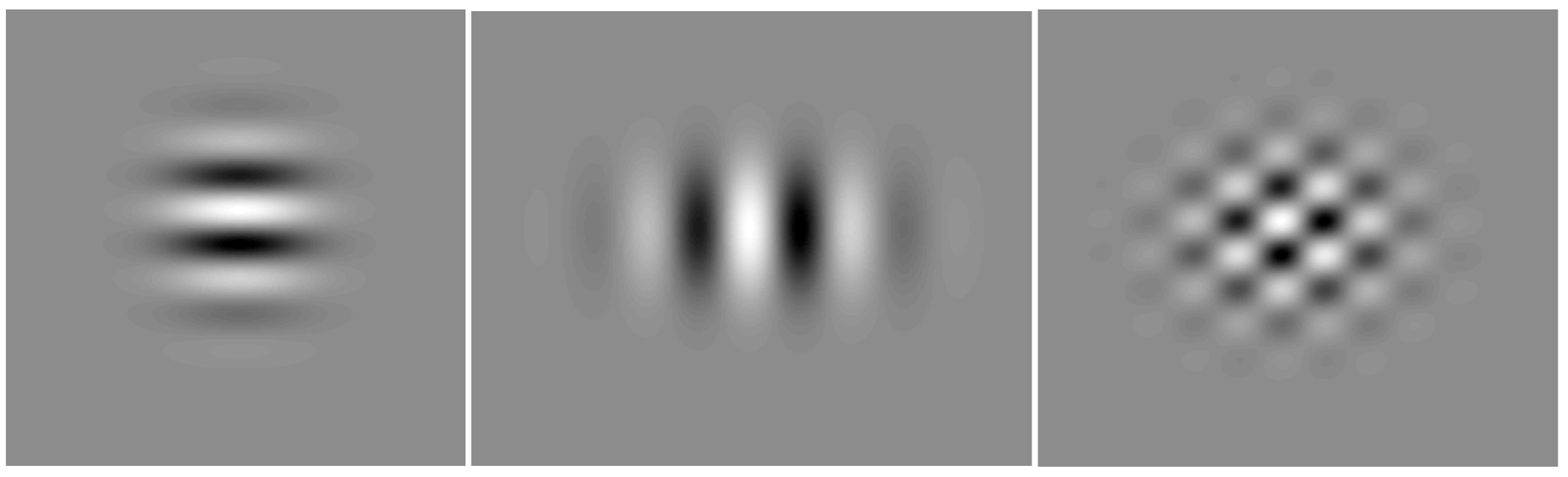}  
\caption{Wavelets associated with the separable basis. The figure shows the LH, HL and HH wavelets in the space domain.}
\label{DWT_checker_effect}
\end{figure}

 	A drawback of $2$D separable wavelets is their preferential response to horizontal and vertical features. Fig. \ref{DWT_checker_effect} shows the three separable  wavelets arising from the separable construction. The pulsation of the LH and HL wavelets are oriented along the directions along which the constituent $1$D wavelets operate. However, the HH wavelet, with its constituent $1$D wavelets operating along orthogonal directions, does not exhibit orientation purely along one direction; instead it shows a checkerboard appearance with simultaneous pulsation along the diagonal directions. 
	 	 		 		 
	 This is exactly where the analytic wavelet $\psi_a=\psi+j\mathcal{H}\{\psi\}$, with its one-sided frequency spectrum comes to the rescue: if instead of employing separable wavelets of the form $\psi(x)\psi(y)$, complex wavelets of the form $\psi_a(x)\psi_a(y)$ are used, then the corresponding spectrum $\widehat{\psi}_a(\omega_x)\widehat{\psi}_a(\omega_y)$ will have only one passband, and consequently the real wavelets $\mathfrak{Re} (\psi_a(x)\psi_a(y))$ and $\mathfrak{Im} (\psi_a(x)\psi_a(y))$ will indeed be oriented.
	
   The motivation then is to use HT pairs of $1$D biorthogonal wavelets to construct oriented $2$D wavelets. In particular, we do so by appropriately combining four separable biorthogonal wavelet bases using Theorem \eqref{prop}. To begin with, we immediately identify the two scaling functions $\varphi_p(x)=\varphi(x)$ and $\varphi_q(x)=\varphi'(x)$, associated with the analytic wavelet $\psi_a(x)=\psi(x)+j\psi'(x)$, where $\psi'(x)=\mathcal{H}\psi(x)$. This naturally leads to the possibility of four separable biorthogonal wavelet bases corresponding to the following possible choices of approximation spaces:  $V(\varphi)\otimes V(\varphi), V(\varphi) \otimes V(\varphi'),V(\varphi') \otimes V(\varphi)$ and $V(\varphi') \otimes V(\varphi')$. In fact, as will be demonstrated shortly, we will employ all of these to obtain a balanced construction.

	First, we identify the separable wavelets corresponding to the four scaling spaces:
\begin{alignat}{2}
\label{b}
\psi_{1}(\x)&=\varphi(x)\psi(y), &  \hspace{6mm}  \psi_{4}(\x)&=\varphi(x)\psi'(y),  \nonumber \\
\psi_{2}(\x)&=\psi(x)\varphi(y),  & \hspace{6mm}  \psi_{5}(\x)&=\psi(x)\varphi'(y),  \nonumber \\
\psi_{3}(\x)&=\psi(x)\psi(y), &  \hspace{6mm}  \psi_{6}(\x)&=\psi(x)\psi'(y),  \nonumber \\  \nonumber \\
{\psi}_{7}(\x)&=\varphi'(x)\psi(y),  & \hspace{6mm} {\psi}_{10}(\x)&=\varphi'(x)\psi'(y), \nonumber \\ 
{\psi}_{8}(\x)&=\psi'(x)\varphi(y), & \hspace{6mm} {\psi}_{11}(\x)&=\psi'(x)\varphi'(y), \nonumber \\ 
{\psi}_{9}(\x)&=\psi'(x)\psi(y), & \hspace{6mm}  {\psi}_{12}(\x)&=\psi'(x)\psi'(y). 
\end{alignat}
The corresponding dual wavelets $\tilde \psi_{\ell}(\x)$ are specified identically except that the dual wavelets are used instead of the primal ones. Finally, by judiciously using the one-sided spectrum of the analytic wavelet $\psi_a(x)=\psi(x)+j\psi'(x)$, and by combining the four separable wavelet bases \eqref{b}, we arrive at the following wavelet specifications:
\begin{equation}
\begin{split}
\label{def_CW}
\Psi_{1}(\x) &= \psi_a(x) \varphi(y)=\psi_{2}(\x)+j\psi_{8}(\x) ,   \\
\Psi_{2}(\x) &= \psi_a(x)  \varphi'(y)=\psi_{5}(\x)+j\psi_{11}(\x),  \\
\Psi_{3}(\x) &= \varphi(x) \psi_a(y)=\psi_{1}(\x)+j\psi_{4}(\x),    \\
\Psi_{4}(\x) &= \varphi'(x) \psi_a(y)=\psi_{7}(\x)+j\psi_{10}(\x),  \\
\Psi_{5}(\x) &= \frac{1}{\sqrt{2}}\psi_a(x)  \psi_a(y)  \\
&=\left(\frac{\psi_{3}(\x)-\psi_{12}(\x)}{\sqrt{2}}\right)+j\left(\frac{\psi_{6}(\x)+\psi_{9}(\x)}{\sqrt{2}}\right),  \\
\Psi_{6}(\x) &= \frac{1}{\sqrt{2}}\psi^{\ast}_a(x)  \psi_a(y) \\
&=\left(\frac{\psi_{3}(\x)+\psi_{12}(\x)}{\sqrt{2}}\right)+j\left(\frac{\psi_{6}(\x)-\psi_{9}(\x)}{\sqrt{2}}\right).
\end{split}
\end{equation}

	The dual complex wavelets, $\tilde \Psi_k(\x)$, are specified in an identical fashion using the dual wavelets $\tilde \psi_{\ell}(\x)$. Importantly, the above construction is \textit{complete} in the sense that it involves all the $4\times 3=12$ separable wavelets of the four parallel multiresolutions. The factor $1/\sqrt{2}$ ensures normalization: the real and imaginary components of the six complex wavelets have the same norm. 

\subsection{Directional Selectivity and Shift-Invariance}
\label{HT_2D}

	A real wavelet has a bandpass spectrum that is symmetric w.r.t. to the origin.  As a result, for the wavelet to be oriented, it is necessary that its spectrum be bandpass only along one preferential direction. We claim that the real and imaginary components of the above complex wavelets are oriented along the primal directions $\theta_1=\theta_2=0$, $\theta_3=\theta_4=\pi/2, \theta_5=\pi/4$, and $\theta_6=3\pi/4$, respectively. Indeed, it is easily seen that the support of $\widehat{\Psi}_1(\bw)$ and $\widehat{\Psi}_2(\bw)$ is restricted to the half-plane $\{(\omega_x ,\omega_y ): \omega_x > 0\}$, since their Fourier transform can be written as $\widehat{\Psi}_k(\bw)=\left(1+\mathrm{sign}(\omega_x)\right)\widehat{\mathfrak{Re}(\Psi_k)}(\bw)$. As it is necessary for the real functions $\mathfrak{Re}(\Psi_k)$ and $\mathfrak{Im}(\Psi_k)$ to have symmetric passbands, the claim about their orientation along the horizontal direction then follows immediately. The orientation of the components of the wavelets $\Psi_3(\x)$ and $\Psi_4(\x)$ along the vertical direction follows from a similar argument. 
       
  As far as the wavelet $\Psi_5(\x)$ is concerned, note that $\widehat{\Psi}_5(\bw)=\left(1+\mathrm{sign}(\omega_x)\right)\left(1+\mathrm{sign}(\omega_y)\right) \widehat{\psi}(\omega_x)\widehat{\psi}(\omega_y)$. As a consequence, the support of $\widehat{\Psi}_5(\bw)$ is restricted to the quadrant $\{(\omega_x ,\omega_y ): \omega_x > 0, \omega_y > 0\}$. The symmetry requirements on the spectrums of $\mathfrak{Re}(\Psi_5)(\x)$ and $\mathfrak{Im}(\Psi_5)(\x)$ then establish their orientation along $\pi/4$. A similar argument establishes the orientation of the real components $\mathfrak{Re}(\Psi_6)(\x)$ and $\mathfrak{Im}(\Psi_6)(\x)$ along $3\pi/4$. 
	
	The above-mentioned directional properties allude to some kind of analytic characterization of the complex wavelets. Indeed, akin to the $1$D counterpart, it turns out that the components of the above complex wavelets can also be related via a multi-dimensional extension of the HT that provides further insights into the orientations of the wavelets. In particular, we consider the following directional version of the HT \cite{partialHilbert}:
\begin{equation}
\label{directionalHT}
\mathcal{H}_{\theta}f(\x) \stackrel{\mathcal{F}}{\longleftrightarrow} -j\mathrm{sign}(\bw^T\u_{\theta})\hat{f}(\bw),
\end{equation}
specified by the unit vector $\u_{\theta}=(\cos \theta,\sin \theta)$ pointing in the direction $0 \leqslant \theta < \pi$. That is, the directional HT is performed with respect to the half-spaces $\{\bw: \bw^T\u_{\theta} > 0\}$ and $\{\bw: \bw^T\u_{\theta} < 0\}$ specified by the vector $\u_{\theta}$, and it maps the directional cosine $\cos(\bw_{\theta}^T\x)$ into the directional sine $\sin(\bw_{\theta}^T\x)$. Based on the wavelet definitions \eqref{def_CW}, the following correspondences (for a proof see \S\ref{A4}) can then be derived:
\vspace{0.05in}
\begin{proposition} 
\label{prop_HT2}
The real and imaginary components of the complex wavelets $\Psi_k(\x)$ form directional HT pairs. In particular:
\begin{align}
\label{HT2}
\mathfrak{Im}(\Psi_k(\x))&= \mathcal{H}_{\theta_k} \mathfrak{Re}(\Psi_k(\x)), \ 1 \leqslant  k \leqslant 6. 
\end{align} 
\end{proposition} 
 \vspace{0.05in}
 
 	A significant problem with the decimated DWT is that the critical down-sampling makes it shift-variant. The redundancy of the dual-tree transform has been successfully exploited for partially mitigating this shift-variance problem \cite{kingsbury1,kingsbury2}. Our design further mitigates this shift-variance problem by using a finer sub-sampling scheme in the $0^\circ$ and $90^\circ$ directions. Observe that $\Psi_1(\x) \approx \Psi_2(x,y-1/2)$ owing to the fact that $\beta_{\tau}^{\alpha}(x) \approx \beta_{\tau+1/2}^{\alpha}(x+1/2)$. These wavelets provide a finer sampling in the $y$-direction. Similarly, the vertical wavelets give us a finer sampling in the $x$-direction. 

\begin{figure*}
\centering
\includegraphics[width=100mm,height=50mm]{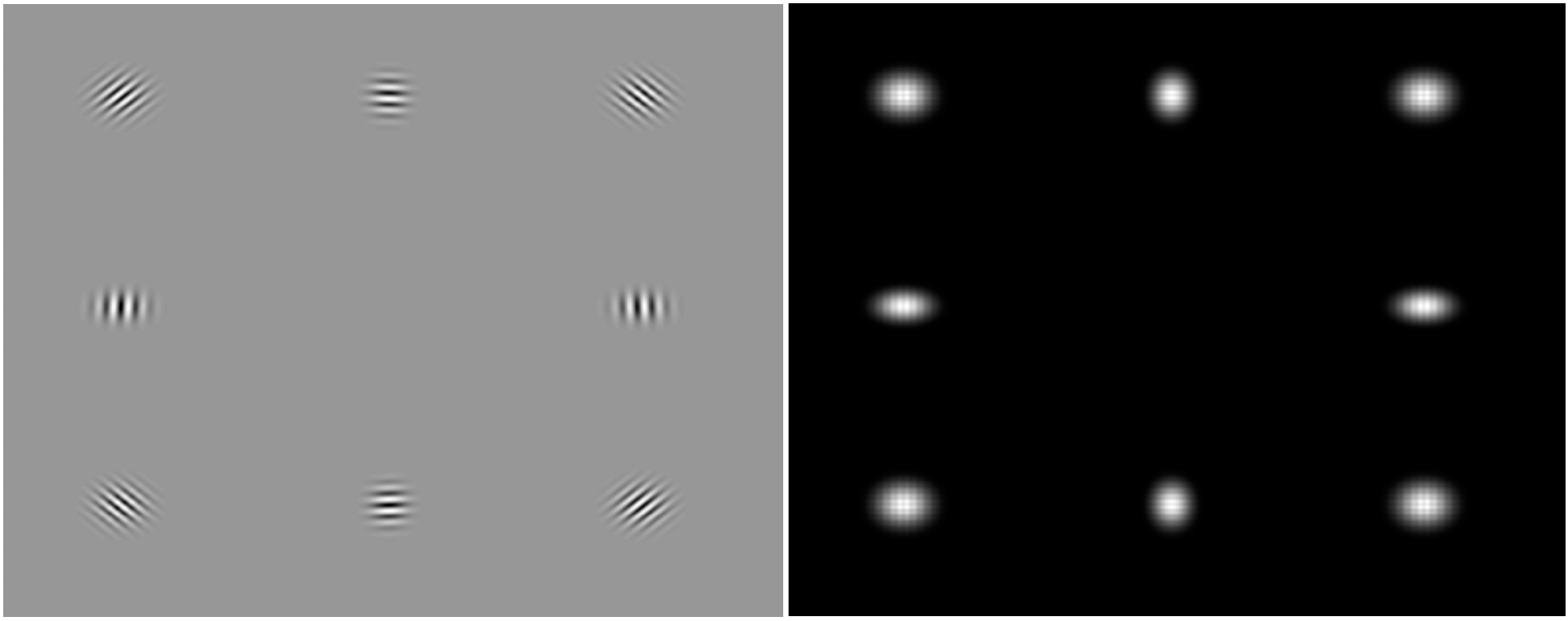}  
\caption{$2$D Gabor-like Wavelets. Left: Real component of the six complex wavelets, Right:  Magnitude envelope of the six complex wavelets. The diagonally placed wavelets are identical, they are used twice to balance the representation.}
\label{gabor_wavelets}
\end{figure*} 

\subsection{Gabor-like Wavelets}
\label{2Dgabor}

	Daugman generalized the Gabor function to the following $2$D form 
\begin{equation}
\label{gabor}
\mathscr{G}(\x)=\frac{1}{2\pi \sigma_x \sigma_y} \Exp^{-\Big(\frac{(x-x_0)^2}{2\sigma_x^2}+\frac{(y-y_0)^2}{2\sigma_y^2}\Big)}\Exp^{j(\xi_0 x+ \nu_0 y)}, 
\end{equation}
involving the modulation of an elliptic Gaussian using a directional plane-wave, to model the receptive fields of the orientation-selective simple cells in the visual cortex  \cite{Daugman}.    

	We are particularly interested in the dual-tree wavelets, denoted by $\mathscr{G}_1(\x; \alpha, \tau),\ldots,\mathscr{G}_6(\x; \alpha, \tau)$, derived from the quadrature B-spline wavelets $\psi_p(x)=\psi^{\alpha}_{\tau}(x)$ and $\psi_q(x)=\psi^{\alpha}_{\tau+1/2}(x)$. These complex wavelets inherit the asymptotic properties of the constituent spline functions. Indeed, by appropriately modifying the proof in \cite{convergence}, it can be shown that  
\begin{equation}
\label{gaussian_conv}
\beta^{\alpha}_{\tau}(x) \sim  \frac{1}{\sqrt{2\pi} \sigma}  \exp\left(-\frac{(x-\tau)^2}{2 \sigma^2}\right)
\end{equation}
for sufficiently large $\alpha$, where $\sigma=\sqrt{\alpha+1}/2\sqrt 3$. This, combined with \eqref{asymp_formula}, then results in the following asymptotic characterization: 
\vspace{0.05in}
\begin{proposition}(\emph{$2$D Gabor-like Wavelets})
\label{expression_CW} 
The complex wavelets $\mathscr{G}_k(\x;\alpha, \tau)$ resemble the $2$D Gabor functions for sufficiently large $\alpha$: 
\begin{equation}
\begin{split}
\mathscr{G}_{1}(\x;\alpha, \tau)&\sim M_1 \Exp^{-\left(\frac{(x-1/2)^2}{\sigma_1^2}+\frac{(y-\tau)^2}{\sigma_2^2}\right)} \Exp^{j(\omega_0 x-\frac{\omega_0}{2}-\pi \tau)}  , \\
\mathscr{G}_{2}(\x;\alpha, \tau)&\sim M_1 \Exp^{-\left(\frac{(x-1/2)^2}{\sigma_1^2}+\frac{(y-\tau-1/2)^2}{\sigma_2^2}\right)} \Exp^{j(\omega_0 x-\frac{\omega_0}{2}-\pi \tau)}  ,\\
\mathscr{G}_{3}(\x;\alpha, \tau)& \sim M_1 \Exp^{-\left(\frac{(x-\tau)^2}{\sigma_2^2}+\frac{(y-1/2)^2}{\sigma_1^2}\right)} \Exp^{j(\omega_0 y-\frac{\omega_0}{2}-\pi \tau)} , \\
\mathscr{G}_{4}(\x;\alpha, \tau)&\sim M_1 \Exp^{-\left(\frac{(x-\tau-1/2)^2}{\sigma_2^2}+\frac{(y-1/2)^2}{\sigma_1^2}\right)} \Exp^{j(\omega_0 y-\frac{\omega_0}{2}-\pi \tau)} ,\\
\mathscr{G}_{5}(\x;\alpha, \tau) &\sim  M_2 \Exp^{-\left(\frac{(x-1/2)^2}{\sigma_1^2}+\frac{(y-1/2)^2}{\sigma_1^2}\right)} \Exp^{j(\omega_0(x+y)-\omega_0-2\pi\tau)} , \\
\mathscr{G}_{6}(\x;\alpha, \tau) &\sim  M_2 \Exp^{-\left(\frac{(x-1/2)^2}{\sigma_1^2}+\frac{(y-1/2)^2}{\sigma_1^2}\right)}\Exp^{j\omega_0(y-x)},
\end{split}
\end{equation}
where $M_1=2\sqrt{3} M_0^{\alpha+1}\Delta\omega_0/\pi(\alpha+1)$;  $M_2= 2M_0^{2(\alpha+1)} \Delta\omega^2_0/\pi(\alpha+1)$; $\sigma_1=\sqrt{\alpha+1}/\Delta\omega_0$; and $\sigma_2=\sqrt{(\alpha+1)/6}$.\\
\end{proposition}
\vspace{0.05in}

 	We call the wavelets ``Gabor-like'' since they form approximates of $2$D Gabor functions similar to the ones proposed by Daugman \eqref{gabor}. The dual-tree transform (cf. $\S$\ref{VIII}) corresponding to a specific family of such Gabor-like wavelets (fixed $\alpha$ and $\tau$) results in a multiresolution, directional analysis of the input image $f(\x)$ in terms of the sequence of projections $\langle f, 2^i \mathscr{G}_k(2^i \x-\m; \alpha, \tau) \rangle$. Fig. \ref{gabor_wavelets} shows the $2$D Gabor wavelets corresponding to $\alpha=6$ and $\tau=0$. The ensemble shows the modulus $|\mathscr{G}_k(\x; 6,0)|$ and the real component $\mathfrak{Re}(\mathscr{G}_k(\x;6,0))$ of the six complex wavelets; the former shows the pulsations of the directional plane waves, whereas the latter shows the elliptical Gaussian envelopes. 

\subsection{Discussion}	

	Before moving on to the implementation, we digress briefly to discuss certain key aspects of our construction:  

\begin{itemize}
\item \emph{Directionality}: The six complex wavelets in Kingsbury's DT-$\mathbb{C}$WT scheme are oriented along the directions: $\pm 15^{\circ}, \pm 45^{\circ}, \mbox{and} \pm 75^{\circ}$ \cite{kingsbury2}. Though we use similar separable building blocks in our approach, our wavelets are oriented along the four principal directions: $0,\pi/4, \pi/2$ and $3\pi/4$. The added redundancy along the horizontal and vertical directions yields better shift-invariance along these directions. Alternatively, we could also have applied Kinsbury's construction to obtain Gabor-like wavelets orientated along $\pm 15^{\circ}, \pm 45^{\circ}$, and $\pm 75^{\circ}$.

\item  \emph{Localization Vs. Frame Bounds}: In this paper, we placed emphasis on time-frequency localization, and were able to construct new wavelets that converge to Gabor-like functions. These basis functions should prove useful for image analysis tasks such as extraction of AM-FM information and texture analysis. However, the price to pay for this improved localization is that the associated transform---in contrast with the transforms constructed by Kingsbury \textit{et al.} \cite{CTDWT,kingsbury_ortho}---is no longer tight, and consequently requires a different set of reconstruction filters. Nevertheless, the tightness of the frame-bounds---a desirable property for image processing applications such as denoising and compression---can, in principle, also be achieved within our proposed framework by replacing the B-spline wavelets with the orthonormal ones (Battle-Lémarie  wavelets).

\item \emph{Analytic Properties}:  Our method of construction takes a primary wavelet transform and obtains an exact HT pair using a simple unitary mapping. The consequence is that all fundamental approximation-theoretic properties of continuous-domain wavelets, such as vanishing moments and regularity, are automatically preserved, and that the associated filters inherit  an exact one-sided response. We also obtain an explicit space-domain expression for the Gabor-like wavelets.

\item \emph{Multidimensional HT properties}: The directional HT correspondences \eqref{HT2} for our complex wavelets follows as a direct consequence of the tensor-product construction. We would however like to note that there exist other multidimensional extensions of the HT as well: the ``single-orthant'' extension of Hahn \cite{Hahn_paper} involving the boundary distribution of analytic functions; the ``hypercomplex'' extension due to B\"ulow \textit{et al.} \cite{Bulow}; the ``monogenic'' signal due to Felsberg \textit{et al.} \cite{Felsberg}; and the spiral-phase quadrature transform of Larkin \textit{et al.} \cite{Larkin}. The last two in the list are closely related to the Riesz transform of classical harmonic analysis \cite{Stein_Weiss}. Design of directional wavelets based on these and other alternative extensions are a promising topic of research \cite{Quarternion_Baranuik,HyperAnalytic}.
\end{itemize}

\begin{figure*}[htdp]
\centering
\includegraphics[width=0.9\linewidth]{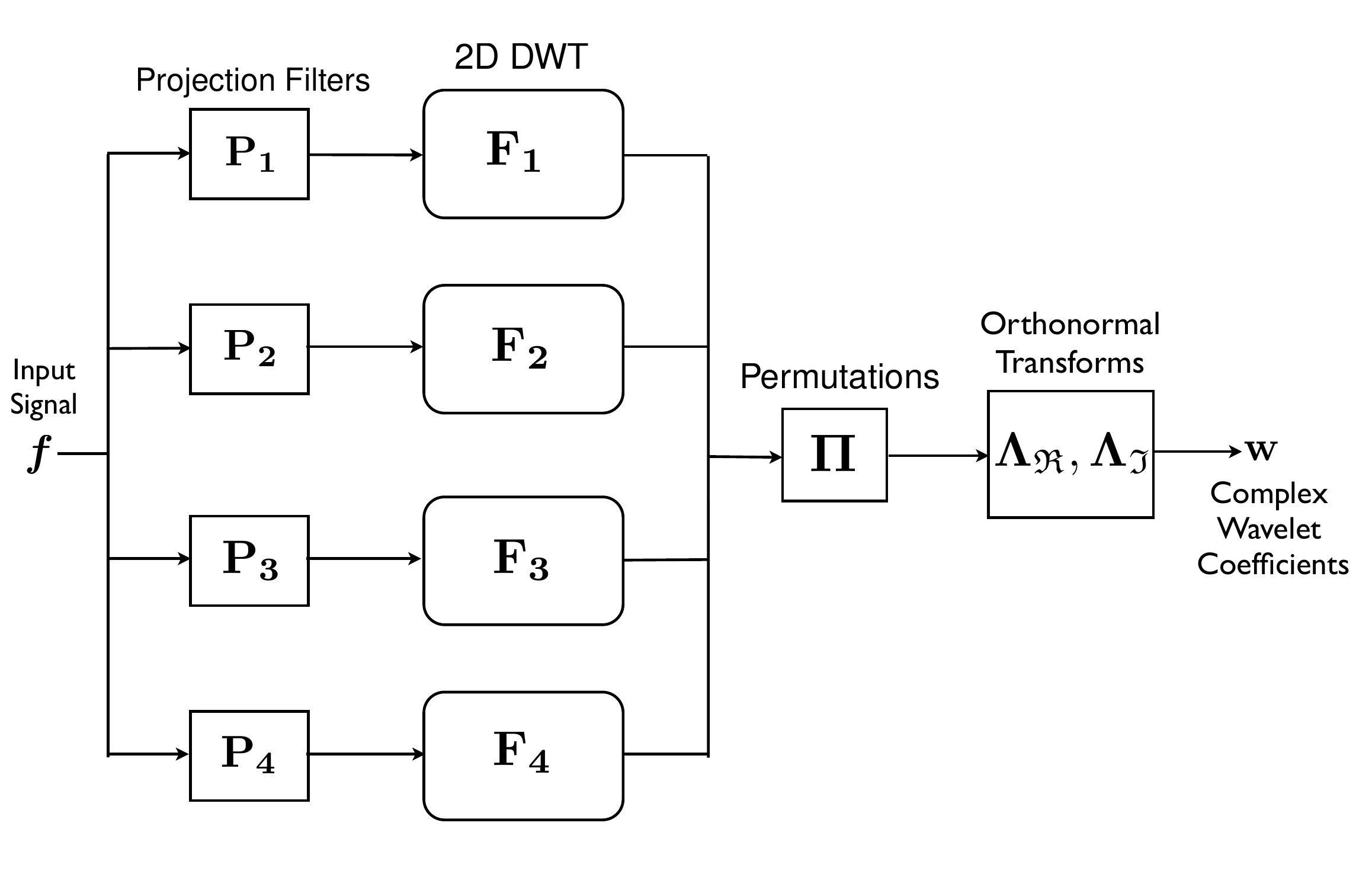}  
\caption{Block Diagram of the $2$D Complex Wavelet Transform.}
\label{CWT}
\end{figure*} 

\section{$2$D IMPLEMENTATION}
\label{VIII}

\textit{Pre-filtering}: The input signal has to be projected onto each of the four separable spaces of the form $V(\varphi_p) \otimes V(\varphi_q)$ before  initiating the multiresolution decompositions. As in the $1$D setting, the orthogonal projection is achieved in a separable fashion using an appropriate pre-filter along each dimension. In particular, if $\{f[\k]\}_{\k \in \mbb{Z}^2}$ be the uniform samples of a bandlimited input signal $f(\x)$, then the projection coefficients are given by $c_0^{LL}[\k]=(f \ast p)[\k]$, where the separable pre-filter $p[\k]$ is specified by $\sum p [k_1,k_2] \Exp^{-j(k_1\omega_x+k_2\omega_y)}=\widehat{\dual\varphi}_{\mathfrak{r}}(\omega_x) \widehat{\dual\varphi}_{\mathfrak{s}} (\omega_y)$ for $\bw=(\omega_x,\omega_y)$ in $(-\pi,\pi)^2$. 

	In general, there would be four such projections $c_0^{LL}(n)=f \ast p_{n}, 1\leqslant n \leqslant 4$, corresponding to the $2$D pre-filters $p_1,\ldots,p_4$ associated with the four approximation spaces. Note that the filters can be implemented efficiently through successive $1$D filtering along either dimension. \\

\begin{figure}[htdp] 
\centering
\includegraphics[width=1.0\linewidth]{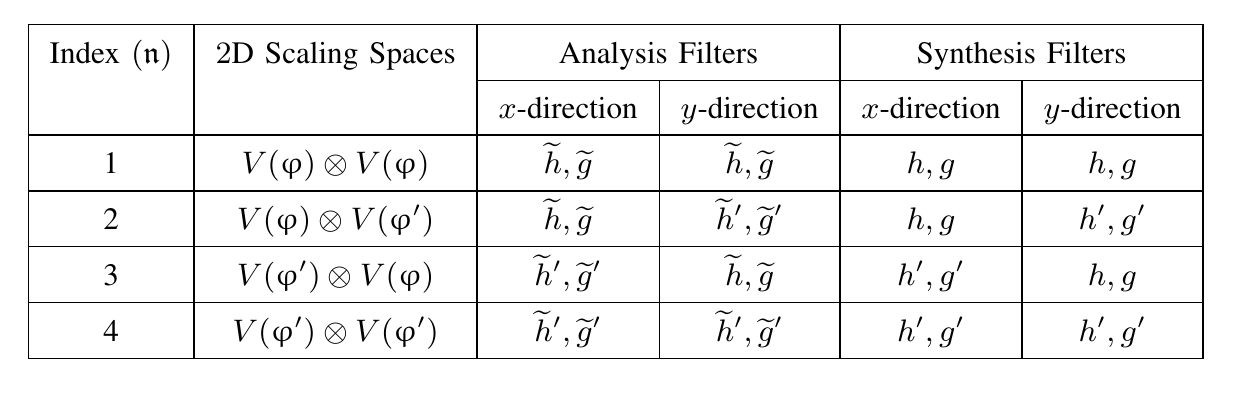}  
\caption{Analysis and synthesis filters corresponding to the four multiresolutions.}
\label{2Dfilters}
\end{figure}

\textit{Analysis}: We consider the implementation aspects for a finite input signal $\f \in \mbb{R}^{M \times N}$. The transform, corresponding to the complex wavelets \eqref{def_CW}, involves four separable DWTs with different filters applied along the $x$ and $y$ directions (cf. Table \ref{2Dfilters} for the list of filters), and result in four subbands at each decomposition level. Specifically, let $\c^{LL}_i(n),\c^{LH}_i(n),\c^{HL}_i(n)$ and $\c^{HH}_i(n), 1\leqslant n \leqslant 4,$ denote the low-low, low-high, high-low and high-high subbands, respectively, of the four DWT decompositions at resolution $i=1,\ldots,J$. The low-low subbands $\c^{LL}_0(n)$ are identified as the four set of pre-filtered signals $\c_0^{LL}(n)=\P_{n} \f$, with $\P_{n}$ being the (block) circulant matrices associated with the $2$D pre-filters. The coarser  subbands at levels $i=1,\ldots,J$ are then given by
\begin{eqnarray}
\label{transform}
\c^{LL}_i(n)&=&\F_{n}(\tilde h_x,\tilde h_y) \c^{LL}_{i-1}(n)  \nonumber \\
\c^{LH}_i(n)&=&\F_{n}(\tilde h_x,\tilde g_y) \c^{LL}_{i-1}(n) \nonumber \\
\c^{HL}_i(n)&=&\F_{n}(\tilde g_x,\tilde h_y) \c^{LL}_{i-1}(n) \nonumber \\
\c^{HH}_i(n)&=&\F_{n}(\tilde g_x,\tilde g_y) \c^{LL}_{i-1}(n),
\end{eqnarray}
where $\F_{n}(q_x, q_y)$ denotes the composition of the $n$th DWT matrix (employing analysis filters $q_x$ and $q_y$ in the $x$-direction and $y$-direction), and the downsampling matrix.  

	The complex subbands $\w_i=(\w^1_i, \ldots,\w^6_i), 1\leqslant i \leqslant J,$ are specified by $\w_i=\boldsymbol{\Lambda}_{\mathfrak{R}} \boldsymbol{\zeta}_i+ j \boldsymbol{\Lambda}_{\mathfrak{I}} \boldsymbol{\xi}_i,$ where 
\begin{equation}
\label{grouping}
\begin{split}
\boldsymbol{\zeta}_i= \big(\c^{HL}_i(1), \c^{HL}_i(2), \c^{LH}_i(1), \c^{LH}_i (3),  \c^{HH}_i(1), \c^{HH}_i(4)\big), \\
\boldsymbol{\xi}_i= \big( \c^{HL}_i(3), \c^{HL}_i(4), \c^{LH}_i(2), \c^{LH}_i(4), \c^{HH}_i(2), \c^{HH}_i(3)\big),
\end{split}
\end{equation}
are obtained through a particular permutation of the $12$ highpass subbands; and the block matrices $\boldsymbol{\Lambda}_{\mathfrak{R}}$ and $\boldsymbol{\Lambda}_{\mathfrak{I}}$ are specified as
\begin{equation}
\label{block_transforms}
\begin{split}
\boldsymbol{\Lambda}_{\mathfrak{R}}=\frac{1}{\sqrt{2}} \left(\begin{array}{cccccc}\sqrt{2}I & 0 & 0 & 0 & 0 & 0 \\0 & \sqrt{2} I & 0 & 0 & 0 & 0 \\0 & 0 & \sqrt{2} I & 0 & 0 & 0 \\0 & 0 & 0 & \sqrt{2} I & 0 & 0 \\0 & 0 & 0 & 0 & I & -I \\0 & 0 & 0 & 0 & I & I\end{array}\right), \\
\boldsymbol{\Lambda}_{\mathfrak{I}}=\frac{1}{\sqrt{2}} \left(\begin{array}{cccccc}\sqrt{2} I & 0 & 0 & 0 & 0 & 0 \\0 & \sqrt{2} I & 0 & 0 & 0 & 0 \\0 & 0 & \sqrt{2} I & 0 & 0 & 0 \\0 & 0 & 0 &  \sqrt{2}I & 0 & 0 \\0 & 0 & 0 & 0 & I & I \\0 & 0 & 0 & 0 & I & -I\end{array}\right).
\end{split}
\end{equation}
In short, the transform can be formally summarized via the frame operation 
\begin{equation}
T: \f \mapsto (\c_J (1),\ldots,\c_J (4), \w_1,\ldots,\w_J)
\end{equation}
involving the sequence of transformations: projections $\P_1,\ldots,\P_4$; discrete wavelet transforms $\F_1,\ldots,\F_4$; permutation $\Pi$, and orthonormal transformations $\boldsymbol{\Lambda}_{\mathfrak{R}}$ and $\boldsymbol{\Lambda}_{\mathfrak{I}}$. Figure \ref{CWT} provides a schematic of these sequence of 
transformations.  \\

\begin{figure*}[htdp] 
\centering
\includegraphics[width=1.0\linewidth]{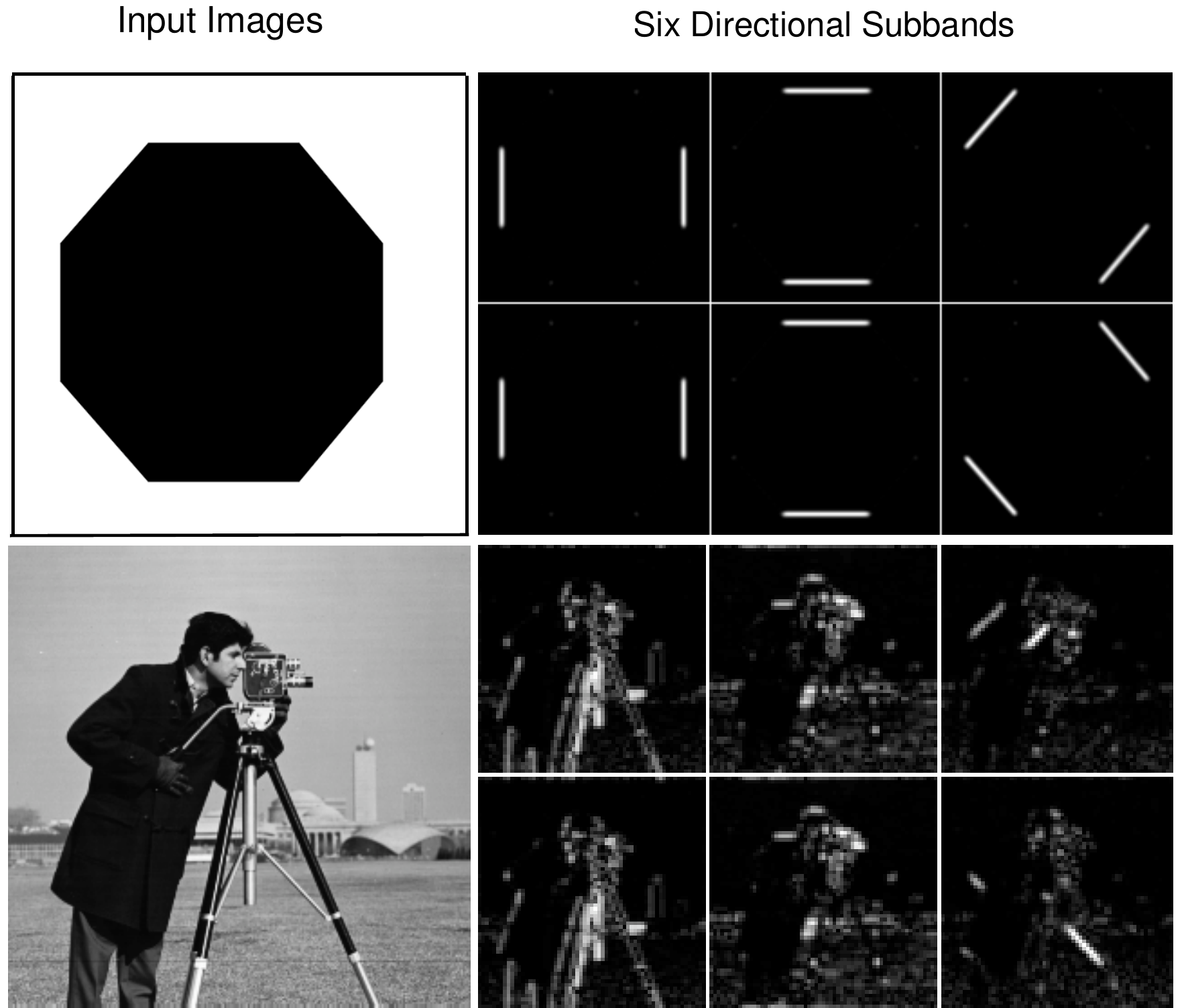}  
\caption{Directional decomposition (one-level) of a synthetic image (Octagon) and a natural image (Cameraman) using the Gabor-like transform. Ordering of the subbands in either case: First column : $|\w_1|, |\w_2|$; Second column $: |\w_3|,|\w_4|$; Third column $:  |\w_5|$ and $|\w_6|$.}
\label{experiment}
\end{figure*} 

\textit{Reconstruction}: Note that the permutation 
\begin{equation}
\label{perm}
\Pi: \{\c^{LH}_i(n),\c^{HL}_i(n),\c^{HH}_i(n)\}_{1 \leqslant n \leqslant  4} \mapsto (\boldsymbol{\zeta}_i,\boldsymbol{\xi}_i)
\end{equation}
involved in \eqref{grouping} is invertible, and that the matrices $\boldsymbol{\Lambda}_{\mathfrak{R}}$ and $\boldsymbol{\Lambda}_{\mathfrak{I}}$ are orthonormal, with corresponding inverses given by $\boldsymbol{\Lambda}^T_{\mathfrak{R}}$ and $\boldsymbol{\Lambda}^T_{\mathfrak{I}}$ respectively.  Starting with the complex wavelet subbands $ \w_1,\ldots, \w_J$, the highpass subbands $\c^b_i(1),\ldots, \c^b_i(4)$ corresponding to the bands $b=HL,LH,$ and $HH$, are then computed from the vectors $\boldsymbol{\zeta}_i=\boldsymbol{\Lambda}_{\mathfrak{R}}^T \mathfrak{Re}(\w_i)$ and $\boldsymbol{\xi}_i=\boldsymbol{\Lambda}_{\mathfrak{I}}^T \mathfrak{Im}(\w_i)$,
via the permutation $\Pi^{-1}$ at levels $i=1,\dots,J$. These, along with the lowpass subbands $\c^{LL}_J(1),\ldots,\c^{LL}_J(4)$, are then used to reconstruct the projected signals $\c_0^{LL}(1),\ldots,\c_0^{LL}(4)$ using the recursion  
\begin{eqnarray}
\label{synthesis}
\c_i^{LL}(n)=\F_{n}(h_x, h_y) \c^{LL}_{i+1} (n)+\F_{n}(h_x,g_y) \c^{LH}_{i+1}(n), \nonumber \\
+\F_{n}(g_x, h_y) \c^{HL}_{i+1}(n) + \F_{n}(g_x,g_y) \c^{HH}_{i+1}(n)
\end{eqnarray}
for $i=J-1,\ldots, 0$. Here, $\F_{n}(m_x, m_y)$ represents the composition of the upsampling matrix and the synthesis matrix corresponding to the $n$th DWT, with filters $m_x$ and $m_y$ in the $x$-direction and $y$-direction, respectively, as specified in Table \ref{2Dfilters}. The input signal samples are finally recovered as $\f=1/4 \sum_{n=1}^4 \P^{-1}_{n}\c_0^{LL}(n)$.

\textit{$2$D Gabor-like Transform}: The Gabor-like transform is based on the analytic B-spline wavelets specified in \S\ref{2Dgabor}, where the complex subbands $\mathcal{\w}^k_i[\m]$ represent the directional decompositions of the input image along the four primal directions using the optimally-localized Gabor-like wavelets $\mathscr{G}_k(\x;\alpha, \tau)$ at different resolutions. The filterbank analysis \eqref{transform} and synthesis \eqref{synthesis} operations are implemented in a separable fashion using the $1$D spline DWT filters specified in \eqref{DWT1} and \eqref{DWT2}. 
	
	Fig. \ref{experiment} shows the magnitude response of the six complex wavelet subands obtained by applying our Gabor-like transform to a synthetic and a natural image. In particular, the wavelet subbands corresponding to the synthetic image, with directional edges along $0,\pi/4,\pi/2$ and $3\pi/4$, highlight the directional-selectivity of the transform. The simulation was carried out in MATLAB $7.5$ on a Macintosh $2.66$\,GHz Intel dual-core system. The average execution time for one-level wavelet analysis and reconstruction (including pre- and post-filtering) of a $512 \times 512$ image is $1.2$ seconds, and the reconstruction error is of the order of $10^{-16}.$

\section{CONCLUDING REMARKS}
\label{IX}

     The primary objective of this contribution was to combine the attractive features of Gabor analyses and multiresolution wavelet transforms into a single theoretical framework, and to provide a fast  algorithm for the same. Specifically, we proposed a formalism for constructing \textit{exact} HT pairs of biorthogonal wavelets based on (i) the B-spline factorization theorem, and (ii) a natural discretization of the continuous HT filter identified via the action of the HT on fractional B-splines. Based on this methodology, analytic wavelets resembling the Gabor function were then designed using HT pair of B-spline wavelets. 
     
    We then extended our scheme to $2$D: starting from HT pair of $1$D biorthogonal wavelet basis, we constructed directional complex wavelets by appropriately combining four separable biorthogonal wavelet bases. In particular, we related the real and imaginary components of the complex wavelets using a directional extension of the HT. The particular family of wavelets constructed using B-splines was shown to resemble the directional Gabor function family proposed by Daugman. Finally, we demonstrated how the discrete Gabor-like transforms could be implemented using fast FFT-based filterbank algorithms.

\section{ACKNOWLEDGEMENTS}

	The authors would like to thank Dr. T. Blu for sharing his research findings (the asymptotic form \eqref{thiery} in particular) on fractional splines, and Dr. P. Thévenaz and Dr. C. S. Seelamantula for proofreading the manuscript.

\section{APPENDIX}        	

\subsection{Proof of Theorem \ref{construction}:}
\label{A1}	

We begin with the following sequence of equivalences 
\begin{eqnarray}
\label{E2}
\begin{split}
\mathcal{H}\psi(x/2) &=\sum_{k \in \mathbb{Z}} g[k]\mathcal{H}\varphi(x-k)\nonumber   \\
&=\sum_{k \in \mathbb{Z}} g[k] (\mathcal{H}\beta_{\tau}^{\alpha} \ast \varphi_0)(x-k) \nonumber   \\
&=\sum_{k \in \mathbb{Z}} g[k]\Bigg(  \sum_{n \in \mathbb{Z}} d[n] (\beta_{\tau+1/2}^{\alpha} \ast \varphi_0)(\cdot- n) \Bigg)(x-k) \nonumber  \\
&=\sum_{m \in \mathbb{Z}} (g \ast d)[m] \varphi'(x-m), 
\end{split}
\end{eqnarray}
based on \eqref{hilbert}, and the linearity, associativity, and commutativity of the underlying convolution operators. The sufficiency part of the theorem then follows immediately: if $g'[k]=(g \ast d)[k]$, then $\psi'(x)=\mathcal{H}\psi(x)$. 
	
	Conversely, let $\psi'(x)=\mathcal{H}\psi(x)$, so that $\sum g'[k] \varphi'(x-k) =\sum (g \ast d)[k] \varphi'(x-k)$. Now, since $\{\varphi'(\cdot-n)\}$ forms a Riesz basis of the subspace $V(\varphi')=\mathrm{span}_{\ell^2}\{\varphi'(\cdot-k)\}_{k \in \mbb{Z}}$, every element in $V(\varphi')$ necessarily has a unique representation. Hence, $g'[k]=(g \ast d)[k]$.


\subsection{Proof of Proposition \ref{prop_main}:}
\label{A2}

The primal scaling functions can be trivially factorized: $\varphi(x)=(\beta_{\tau}^{\alpha} \ast \varphi_0)(x)$ and $\varphi'(x)=(\beta_{\tau+1/2}^{\alpha} \ast \varphi_0)(x)$, where $\varphi_0$ is the Dirac delta distribution. Similarly, the dual scaling functions can be factorized as $\tilde{\varphi}(x)=(\beta_{\tau}^{\alpha} \ast \tilde{\varphi}_0)(x)$ and $\tilde{\varphi}'(x)=(\beta_{\tau+1/2}^{\alpha} \ast \tilde{\varphi}_0)(x)$, where $\tilde{\varphi}_0=\sum q^{\alpha}[k] \delta(\cdot-k)$ with $\sum q^{\alpha}[k] \Exp^{-j\omega k}=1/A^{\alpha}(\mathrm{e}^{j\omega})$. Note that in the latter case we have particularly used the fact that $A^{\alpha}(\mathrm{e}^{j\omega})$, and hence $q^{\alpha}[k]$, are independent of $\tau$. 

	The proposition then follows from Corollary \eqref{corollary} since the wavelet filters satisfy the sufficiency conditions: $\tilde g'[k]=(d \ast \tilde g)[k]$ and $g'[k]=(d \ast g)[k]$, respectively. Indeed, from \eqref{filter_pair} and \eqref{semi_filter}, we have
\begin{equation*}
\begin{split}
G'(\mathrm{e}^{j\omega})&= \mathrm{e}^{j\omega} A^{\alpha}(-\mathrm{e}^{j\omega})H^{\alpha}_{\tau+1/2}(-\Exp^{-j\omega}) \\
&= \mathrm{e}^{j\omega} A^{\alpha}(-\mathrm{e}^{j\omega}) D(\mathrm{e}^{j\omega}) H_{\tau}^{\alpha}(-\Exp^{-j\omega})  \\
&= D(\mathrm{e}^{j\omega})  G(\mathrm{e}^{j\omega}).
\end{split}
\end{equation*}
The other condition $\tilde G'(\mathrm{e}^{j\omega})=D(\mathrm{e}^{j\omega}) \tilde G(\mathrm{e}^{j\omega})$ can be similarly derived.


\subsection{Derivation of Equation \eqref{prefilter_formula}:}
\label{A3}

	It is well-known that the least-square approximation operator $P_{V(\varphi)}: \mathrm{L}^2(\mbb{R}) \rightarrow V(\varphi)$, defined by
\begin{equation} 
P_{V(\varphi)}f= \arg \min_{f_0 \in V(\varphi)} \|f-f_0\|
 \end{equation}
gives the orthogonal projection of $f(x)$ onto $V(\varphi)$. The solution to the above problem is explicitly given by $P_{V(\varphi)}f(x)= \sum c_0[k]\varphi(x-k)$ where the coefficients are specified by $c_0[k]=\langle f, \dual \varphi(\cdot-k)\rangle$. Here $\dual\varphi(x)$ denotes the dual of $\varphi(x)$ that satisfies the biorthogonality criterion $\langle \varphi,\dual\varphi(\cdot-n) \rangle=\delta[n]$. Moreover, under the constraint that $\dual\varphi(x) \in V(\varphi)$, we recover a unique dual that is specified by the Fourier transform $\widehat{\dual\varphi}(\omega)=\widehat{\varphi}(\omega)/\sum |\widehat{\varphi}(\omega+2\pi k)|^2$ \cite{sampling_paper}. 
 
	Next, using the Poisson summation formula, we derive the expression $C_0(\mathrm{e}^{j\omega})=\sum_{n \in \mathbb{Z}} \widehat {(f \ast \dual\varphi^T)} (\omega+2\pi n)$ for the (discrete) Fourier transform of $c_0[k]$. The bandlimited model $f(x)=\sum f[k] \ \mathrm{sinc} (x-k)$ finally results in the simplification
\begin{eqnarray}
\label{coeff}
C_0(\mathrm{e}^{j\omega})&=& \sum_{n \in \mbb{Z}} \widehat{(f \ast {\dual\varphi}^T)}(\omega+2\pi n) \nonumber \\
&=&F(\mathrm{e}^{j\omega}) \sum_{n \in \mathbb{Z}} \mathrm{rect}\left(\frac{\omega+2\pi n}{2\pi}\right)\widehat{\dual\varphi}(\omega+2\pi n) \nonumber \\
&=&F(\mathrm{e}^{j\omega}) P(\mathrm{e}^{j\omega}),
\end{eqnarray}
where $P(\mathrm{e}^{j\omega})$ equals $\widehat{\dual\varphi}(\omega)$ on $(-\pi,\pi)$, and $F(\mathrm{e}^{j\omega})$ is the Fourier transform of $f[k]$. 


\subsection{Proof of Proposition \ref{prop_HT2}:}	
\label{A4}

 We establish the correspondence for the wavelets $\Psi_1(\x)$ and $\Psi_5(\x)$ (the rest can be derived similarly). The correspondence for the former is direct: $\mathcal{H}_{0} \mathfrak{Re}(\Psi_1(\x))=\mathcal{H}_{0} \{\psi(x)\} \varphi(y)= \psi'(x) \varphi(y)=\mathfrak{Im}(\Psi_1(\x))$. 
 
 Next, note that the Fourier transforms of $\mathfrak{Re}(\Psi_5(\x))$ and $\mathfrak{Im}(\Psi_5(\x))$ can be written as
\begin{align}
\label{keyrelation}
\widehat{\mathfrak{Re}(\Psi_5)}(\bw)&=\frac{1}{\sqrt 2} \big(1+\mathrm{sign}(\omega_x)\mathrm{sign}(\omega_y)\big) \hat\psi(\omega_x) \hat\psi(\omega_y),  \ \mbox{and}\nonumber \\
\widehat{\mathfrak{Im}(\Psi_5)}(\bw)&=-\frac{j}{\sqrt 2} (\mathrm{sign}\big(\omega_x)+\mathrm{sign}(\omega_y)\big) \hat\psi(\omega_x) \hat\psi(\omega_y).
\end{align}
The correspondence $\mathfrak{Im}(\Psi_5(\x))= \mathcal{H}_{\pi/4} \mathfrak{Re}(\Psi_5(\x))$ then follows from the identity $(\mathrm{sign}(\omega_x)+\mathrm{sign}(\omega_y))=\mathrm{sign}(\omega_x+\omega_y) \big(1+\mathrm{sign}(\omega_x)\mathrm{sign}(\omega_y)\big)$.

\bibliographystyle{amsplain}
\bibliography{bibliography.bib}	

\end{document}